# Advancing DeFi Analytics: Efficiency Analysis with Decentralized Exchanges Comparison Service

Evgenii Onishchuk[*]
*Data Scientist*
evg.oni@proton.me

Maksim Dubovitskii[*]
*Analytics Team Lead*
maksmorrow@gmail.com

Eduard Horch[*]
*Senior Data Scientist*
eduardhorch@proton.me

**Abstract**

This empirical study presents the Decentralized Exchanges Comparison Service (DECS), a novel tool developed by 1inch Analytics to assess exchange efficiency in decentralized finance. The DECS utilizes swap transaction monitoring and simulation techniques to provide unbiased comparisons of swap rates across various DEXes and aggregators. Analysis of almost 1.2 million transactions across multiple blockchain networks demonstrates that both 1inch Classic and 1inch Fusion consistently outperform competitors. These findings not only validate 1inch's superior rates but also provide valuable insights for continuous protocol optimization and underscore the critical role of data-driven decision-making in advancing DeFi infrastructure.

## I. INTRODUCTION

Decentralized Finance (DeFi) has emerged as a transformative force within the cryptocurrency sector, establishing new paradigms that surpass traditional financial systems. This innovative ecosystem offers significant advantages to end-users, fundamentally enhancing the conventional financial experience through the elimination of central regulation and intermediaries. Consequently, DeFi architectures facilitate reduced transaction costs and mitigate risks associated with censorship and account suspensions.

At the core of DeFi lies the principle of tokenization. While cryptocurrencies such as Bitcoin and Ethereum serve as native currencies within their respective blockchain ecosystems, tokens represent digital assets generated through smart contracts. Notably, ERC-20 tokens have become the cornerstone of decentralized ecosystems, enabling the development of diverse financial products and services. This tokenization framework has laid the foundation for revolutionary concepts including non-fungible tokens (NFTs), decentralized exchanges (DEXs), and yield farming protocols, thereby shaping the trajectory of future financial systems.

The inception of DeFi can be traced to the emergence of decentralized exchanges, which facilitated token trading within automated liquidity pools. The first generation of automated market makers (AMMs) introduced the groundbreaking concept of continuous, 24/7 market accessibility a feature previously unseen in traditional financial markets [4]. These systems operated autonomously, leveraging self-executing algorithms for price determination without reliance on external oracles. However, despite their innovative nature, these early AMMs faced significant challenges, including liquidity fragmentation and capital inefficiency, stemming from the uniform distribution of liquidity across all price ranges, irrespective of market demand.

Subsequent generations of AMMs, exemplified by protocols such as Curve and Uniswap V3, addressed these limitations through the implementation of concentrated liquidity mechanisms. This advancement allowed liquidity providers to allocate capital more efficiently by focusing it within specific price bands, thereby enhancing capital efficiency and mitigating risks associated with slippage and impermanent loss.

While AMMs have primarily influenced liquidity provision and trading, their impact on other DeFi sectors, including lending, borrowing, and yield farming, has been more indirect. Nevertheless, these foundational concepts have catalyzed the extensive evolution of decentralized financial ecosystems and incentivized liquidity provision through yield farming strategies. However, these initial systems encountered significant obstacles, including liquidity disper-

---

sion, elevated slippage, and prolonged transaction durations.

The identification of these shortcomings underscored the necessity for more sophisticated solutions, leading to the development of DEX aggregators such as 1inch. These aggregators address the aforementioned challenges by optimizing routes across multiple DEXs to minimize price impact and enhance gas efficiency, thereby reducing overall trading costs.

1inch, established with the primary objective of improving price efficiency, has consistently pursued this goal throughout its evolution. The platform's aggregation router initially focused on optimizing trading routes to enhance pricing and minimize slippage. Additionally, the routing algorithm strikes a balance between reducing gas costs and strategically spending more on gas when necessary, as the potential gains from a better exchange rate can far outweigh the extra routing expenses. Subsequently, 1inch introduced a limit order protocol, enabling users to specify preferred trade prices. The synthesis of these concepts culminated in the development of 1inch Fusion, which executes optimal rate limit orders through a Dutch auction mechanism to ensure the most favorable rates for users. The key to maximizing the benefits of this auction mechanism lies in the competition among resolvers — just two resolvers are enough to activate this dynamic, driving the rates down and securing the best possible deal for the user.

As the decentralized exchange sector has matured, the proliferation of exchanges and protocols has highlighted the necessity for precise performance comparisons. The dynamic nature of DeFi, characterized by rapidly fluctuating liquidity conditions and market dynamics, has posed significant challenges to conventional swap assessment methodologies. These issues have prompted the development of the **Decentralized Exchanges Comparison Service (DECS)**, a tool designed to evaluate exchange efficiency in near real-time. The DECS aims to overcome the limitations of previous data comparison approaches by providing a more accurate and, crucially, unbiased perspective on exchange performance within the DeFi ecosystem.

## A. Objectives of the paper

The main objective of this paper is to present DECS developed by 1inch Analytical team which provides a near real-time, comprehensive approach to assessing exchange efficiency in decentralized finance. And also demonstrate the performance of 1inch protocols (Classic and Fusion) compared to other major competitors across different blockchain networks (Ethereum, Arbitrum, Binance Smart Chain and Polygon).

# II. PROBLEM STATEMENT

## A. Challenges in Comparing Exchange efficiency in DeFi

The decentralized finance ecosystem presents unique challenges when attempting to compare the efficiency of various exchanges and aggregators. These challenges stem from the inherent nature of blockchain technology, the dynamic market conditions, and the limitations of traditional data analysis methods [8]. This section will address the key issues that complicate the process of accurately assessing and comparing exchange performances in the DeFi space.

### 1. *Dynamic Liquidity States*

One of the fundamental challenges in comparing DeFi exchanges is the constant fluctuation of liquidity states. Unlike traditional financial markets, where liquidity pools remain relatively stable over short periods, DeFi liquidity can change dramatically from one block to the next. This volatility is due to the atomic state of transactions and decentralized nature of these platforms, where anyone can add or remove liquidity from pools at any time, causing rapid shifts in available assets.

Liquidity in DeFi can be broadly categorized into "hot" and "cold" liquidity. Hot liquidity is the portion that resides in AMMs, always available and easily accessible for trades. In contrast, cold liquidity is held in user wallets, private market makers (PMMs), arbitrageurs, centralized exchanges, and other platforms. Although cold liquidity is not immediately visible, it is activated during trading — especially when large orders begin to shift prices. As the price moves, this cold liquidity steps in to participate, helping to fill orders and reducing slippage, thereby improving the overall trade efficiency for the user.

Furthermore, the prevalence of arbitrage opportunities leads traders to quickly exploit price discrepancies across different platforms, resulting in frequent rebalancing of liquidity pools. The popularity of yield farming strategies also contributes to this volatility, as liquidity providers often move their assets between different protocols to maximize returns, causing sudden changes in pool compositions.

Additionally, the use of flash loans in DeFi adds another layer of complexity to liquidity dynamics. Flash loans allow users to borrow large amounts of assets without collateral, provided they repay the loan within the same transaction. These uncollateralized loans can be used for arbitrage, collateral swaps, or self-liquidations, leading to significant and instantaneous shifts in liquidity across multiple protocols. The impact of flash loans can be substantial, temporarily draining liquidity from one pool and flooding another, further complicating the task of comparing exchange efficiency.

These dynamic liquidity states make it challenging to perform consistent comparisons, as the same trade executed mere seconds apart might yield significantly different rates due to changes in the underlying liquidity landscape. It's worth noting that while such dramatic fluctuations are rare for highly liquid tokens, they occur more frequently with less established / lower liquidity tokens.

### 2. *Exclusion of Private Market Makers (PMMs)*

Another challenge is the inability to include all Private Market Makers (PMMs) in route calculations and comparisons. Sometimes PMMs play a crucial role in providing liquidity and improving trade execution in many DeFi protocols. However, their operations are often opaque and not visible on-chain until a transaction is confirmed.

This lack of visibility means that potential routes involving PMMs may be missed in analyses, leading to a slightly incomplete market view. Without complete PMM data, the true depth of liquidity and available trading opportunities cannot be fully assessed. Consequently, excluding PMMs from routing could potentially underestimate the efficiency of aggregators that actively utilize them in trade optimization, as it would reduce the number of available liquidity sources.

At the same time, some PMMs (and possibly any liquidity sources) may offer incentives to third parties to boost their trading volumes. While potentially beneficial for the market makers and their partners, these arrangements could adversely affect the rates available to end users. Such practices introduce additional complexity to fair market comparison, as these off-chain agreements are not readily apparent in on-chain data.

Additionally, it should be noted that in practice:
- At the moment, PMMs account for less than a tenth of the protocol volumes of DEX aggregators.
- A significant part (about 40%) of PMMs, which account for most of the traffic of all PMMs, are still present in the routes of simulated transactions.

All this greatly minimizes or eliminates this problem altogether.

### 3. *Network Congestion and Gas Price Volatility*

Ethereum and other blockchain networks sometimes experience periods of high congestion, which can significantly impact user experiences and trade outcomes. During peak usage, gas prices can spike dramatically, affecting the overall cost-effectiveness of trades. This congestion can lead to delayed execution, with transactions remaining pending for extended periods, resulting in increased slippage and potentially less favorable trade outcomes.

## B. Intent-related comparison challenges

Intent-based trading systems have gained significant traction in the Defi ecosystem. These systems offer advantages: main one being gasless transactions, some systems include MEV protection as a built-in mechanism, etc. However, comparing the efficiency of intent-based systems presents unique challenges that require a specialized methodology.

Among the most notable implementations of intent-based solutions are 1inch Fusion, UniswapX, and CoW*[2]*. While these systems share the common goal of optimizing trade execution, they differ in their underlying mechanisms and degree of decentralization.

1inch Fusion and UniswapX employ decentralized order execution mechanisms, leveraging on-chain auctions to determine the most efficient executor for each trade. In contrast, CoW utilizes a more centralized approach, with order matching and executor selection managed by off-chain solvers, coupled with a fixed fee structure.

The primary challenge in comparing intent-based systems lies in the indeterminacy of order execution timing. Unlike classic swap protocols, where execution follows an imperative approach, intent-based systems take a more declarative approach, allowing for a delay in order fulfillment to maximize the user's effective amount. This temporal variability introduces complexities in establishing a fair basis for comparison.

To compare classics (refer to traditional DEX aggregation protocols), it does not matter which protocol to use as a base for comparison. Since in both cases it is possible to obtain transactions for simulation and estimation (modeling in a specific block, a specific state of the blockchain). For an intent-based swaps, it is impossible to obtain unambiguous transactions for simulation and evaluate the results, since the actual result of an intent exchange depends on many factors (further market movement, behavior of resolvers, price-curve, etc., these factors are listed in more detail in *Section VI*).

---

These challenges collectively highlight the need for a more sophisticated approach to comparing exchange efficiency in the DeFi space. Traditional methods of analysis, which rely on historical, on-chain data, are insufficient to capture the true dynamics of this rapidly evolving ecosystem.

DECS was developed to solve these issues. This system is designed to monitor real-time transactions by tapping directly into the mempool or recently mined transactions. This provides a more current view of market activities.

The DECS uses simulation techniques to calculate the outcomes of trades across different exchanges under identical market conditions. It also considers factors like gas prices and network congestion in its analysis, providing a more holistic view of trade efficiency. By analyzing fresh trades data, the DECS can infer user intents in current liquidity market state.

This approach allows for a more accurate and nuanced comparison of exchange efficiency, taking into account the complex and dynamic nature of the DeFi ecosystem. By providing real-time, comprehensive analysis, the DECS offers valuable insights that can drive improvements in 1inch protocol design and ultimately contribute to a more efficient and user-friendly DeFi landscape.

## III. Decentralized Exchanges Comparison Service Overview

DECS is designed to provide the most objective and accurate comparison of the effectiveness of swap products in DeFi.

To ensure the analysis closely mirrors real-world conditions, the service utilizes mempool, newly mined transactions, and new orders from intent-based solutions as sources for receiving user requests. This allows DECS to capture real user needs at specific times and current market states.

To parse a transaction or intent order of a DeFi product, the service incorporates processing logic for each method of its smart contracts (including methods of DEX routers and intent-based solutions). Through these sources, DECS can identify exchanges across integrated DeFi products, including internal 1inch swap products (across different products and their versions).

When such swaps are detected, the service extracts the necessary swap components: trading pairs and amounts. DECS then creates an equivalent trade request to the 1inch router, and depending on the type of swap identified:

- For swaps via the router - DECS creates an equivalent trade request to this router
- For swaps via intent-based solutions - DECS retrieves the executed order results

Subsequently, the created equivalent exchanges are analyzed (for router swaps using `debug_traceCall`, and for intent-based solutions by obtaining execution results), enabling DECS to examine their execution details thoroughly.

As a result, DECS obtains the actual values of amounts sent and received by the user, as well as the projected gas consumption for router-based exchanges.

This approach allows to calculate the results of exchanges most objectively, unlike theoretical calculations, for example, when analyzing quotes. It takes into account the current conditions on the blockchain, reflects the real needs of users, the real results of the swaps and the dynamic nature of DeFi liquidity.

*Transparency note*

The validity of the results of this study is primarily ensured through the transparent and detailed description of the methodology. The DECS algorithm and analysis process have been thoroughly documented in this paper, providing a clear roadmap for anyone wishing to understand or replicate this approach.

This level of transparency in this methodology serves as the cornerstone of result validation. Any significant deviation from the described process would lead to noticeably different outcomes, making it straightforward to detect potential inconsistencies. For results validation of this analysis, all the raw data accumulated by the DECS for the period analyzed under review is provided *[3]*.

## IV. Methodology

### A. In-Depth Description of the DECS architecture

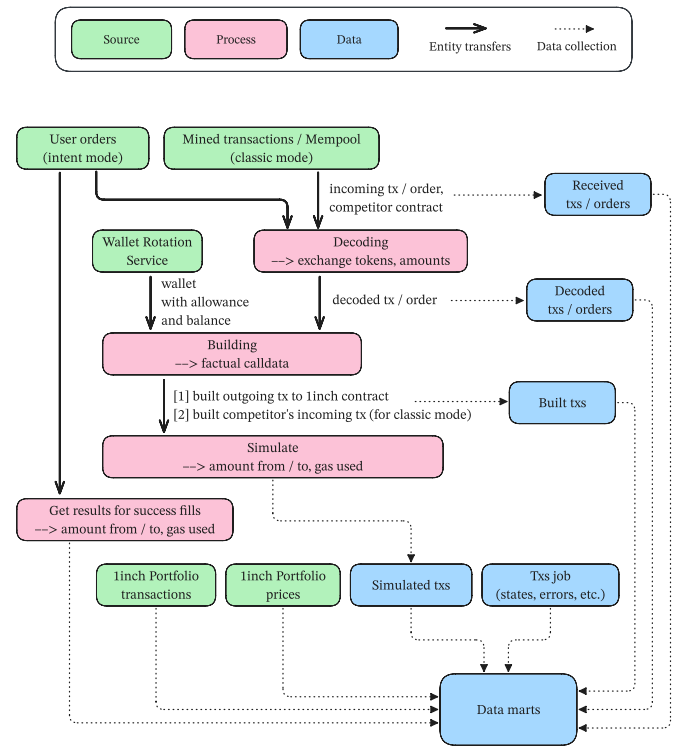

Figure 1: DECS architecture

DECS architecture includes 3 main modules: Getter, Decoder, Builder, Simulator, and overall data handling infrastructure (*Figure 1*).

DECS is built on a modular architecture designed to efficiently process and analyze real-time transaction data. The system comprises several key components that work in concert to provide accurate and timely comparisons of exchange efficiency.

### 1. *The Getter*

The Transaction Getter serves as the initial point of contact with the EVM-compatible networks. It establishes a persistent WebSocket connection to EVM-compatible nodes, allowing it to listen to the mempool in real-time or immediately capture mined transactions.

For each captured transaction, the Getter extracts essential metadata, including:

- The transaction hash
- Sender and recipient addresses
- Transaction value
- Gas price and gas limit
- Input data (calldata)

This raw transaction data is then passed to the Decoder module for further processing.

### 2. *The Decoder*

The Decoder module is responsible for interpreting the raw transaction data and extracting meaningful information about the swap operation. This process involves several steps:

- **Contract Identification:** The Decoder first identifies the specific contract being interacted with based on the recipient address.
- **Function Selector Matching:** It extracts the function selector from the input data and matches it against known selectors for the identified contract.
- **ABI-based Decoding:** Using the appropriate contract ABI (Application Binary Interface), the Decoder parses the input data to extract detailed information about the swap, including:
    - Source token address and amount
    - Destination token address
    - Minimum output amount (if applicable)
    - Deadline or expiry time

The decoded data is then normalized into a standardized format.

### 3. *Wallet Rotation service*

Wallet rotation service interfaces with 1inch indexed blockchain data, to periodically retrieve a set of wallet addresses that are used further in the Builder and Simulator. A set of token approvals, positive token balances and sufficient amount of native token balance are neccessary for wallet selection.

At regular hourly intervals, Wallet Rotation service fetches a fresh batch of wallet addresses. These addresses are then further integrated into the transaction building and simulation processes within the DECS. Constant rotation of wallets is needed in order to ensure successful transaction builds, as the wallet address is a parameter in the building process, and using a wallet with insufficient balance or lacking necessary approvals would result in an error from the node due to insufficient balance or approval.

### 4. *The Builder*

Depending on specific protocol compared, the Transaction Builder can reconstruct equivalent transactions for different exchanges based on the decoded swap parameters. This allows for direct comparisons between the original transaction and potential alternatives on other exchanges or protocols.

Same gas as in incoming transaction is set when building calldata for next steps. Obtained calldata is then passed further with stored routes for aftermath analysis.

### 5. *The Simulator*

The Simulator module is crucial for obtaining the outcomes of both the original transaction and any reconstructed alternatives. It employs several techniques:

- **Debug Trace Call:** The Simulator uses the `debug_traceCall` method to simulate the execution of each transaction on the same block. This provides a detailed trace of the transaction's execution path without actually committing it to the blockchain.
- **Trace Parsing:** The resulting execution trace is parsed to extract:
    - Actual input amount transferred from user
    - Actual output amount transferred to user
    - Gas used
    - Any reverts or errors

### 6. *Data Aggregator and Storage*

The final simulated transaction data, along with the original transaction details and additional helper data (prices, etc.), are stored in a columnar database. Further, this data is transformed and aggregated into a user-friendly data marts.

Having all these modules combined, DECS provides nuanced, meaningful insights into the relative efficiency of various decentralized exchanges and aggregators compared with 1inch protocol.

## B. Intent system comparisons

To address challenges depicted in *Section II.B*, DECS uses a relative comparison methodology. 1inch Classic was designated as the baseline for all comparisons, providing a consistent reference point against which both competitor protocols and 1inch own intent-based solution (1inch Fusion) can be evaluated.

Building upon this baseline, pairwise comparisons are made between 1inch Classic and each intent-based protocol under examination. Currently, two intent solutions are connected to DECS, 1inch Fusion and UniswapX, since:

- Both protocols provide API access to order books and offer comparable functionalities in the intent-based trading space.
- UniswapX employs a decentralized solution, similarly to 1inch Fusion's architecture.

As due to reasons in *Section II.B* it is impossible to directly compare intent-based protocol beetween each other, it is possible to indirectly compare them (*Section V.B.3*) when using 1inch Classic as a baseline.

## C. Postprocessing logic & definitions

### 1. *Data Integration and Enrichment*

Initially, the raw simulation results are ingested into data warehouse. These results are then enriched through integration with two key data sources:

- Mined Transaction Data: This provides essential context about the actual execution of transactions on the blockchain.
- Token Price Data: Proprietary decentralized exchange price engine is used as the primary source for token valuations. In cases where this data is unavailable, spot prices from the 1inch API are used to provide comprehensive coverage. The specific prices source usage is less significant than ensuring that both participants' rates are compared utilizing the same source, as the critical factor is the difference in USD equivalents rather than the USD amounts themselves.

To maintain the integrity and relevance of the analysis, several filtering criteria are applied:

- Exact Output Transactions: excluded transactions where users specify an exact output amount, as this functionality is not supported in the 1inch protocol.
- Price Data Quality: Transactions associated with unreliable or missing price data are removed from the analysis set.

$$\text{mined\_block}^{\dagger} - \text{simulation\_block} \leq X \qquad (1)$$

Where $X$ varies by blockchain from 0 to:

| Blockchain | Maximum | Top cases |
|---|---|---|
| Ethereum | 4 blocks | 1 (57,44%), 0 (24,04%) |
| Binance Smart Chain | 16 blocks | 1 (52,29%), 0 (40,30%) |
| Arbitrum | 192 blocks | 6 (48,69%), 5 (33,33%) |
| Polygon | 24 blocks | 1 (67,18%), 0 (25,67%) |

The (1) constraint ensures the correctness of the simulation: the liquidity state used in build step closely mirrors the actual market conditions at the time of user's intended swap request before submitting it to the network.

At the same time, an important component of the fairness in post-calculations is using suitable gas price (see *Section IV.C.3*).

### 2. *Transaction Types*

DECS distinguishes between two types of transactions.

**Incoming Transactions** refer to the original recently mined or recently detected transactions in the mempool and further built and simulated. These transactions embody the user's initial intent and serve as the baseline against which other potential protocols are compared. In formulas and analysis, incoming transactions are denoted as $X_{\text{in}}$

**Outgoing Transactions**, on the other hand, are the simulated transactions that the DECS creates to compare against the incoming transactions. These hypothetical transactions represent how the same swap would perform on 1inch Classic, using identical input parameters (token pair, amount) as the incoming transaction. In notation of this paper, outgoing transactions are represented as $X_{\text{out}}$

By comparing these two transaction types, it is possible to assess the relative efficiency of 1inch against other competitors. This comparison forms the basis of the performance indicators that will be defined next.

### 3. *Performance Metrics*

To quantify the performance difference between incoming and outgoing transactions, several key metrics are used:

1. Winner Determination. The winner of each comparison is determined using the following logic:

---

† For the comparison flow with intent-based solutions, the mined block = the block of the last order fill.

$$\text{winner} = \begin{cases} \textbf{1inch} & \text{if } A_{\text{eff}_{\text{out}}} - A_{\text{eff}_{\text{in}}} > \varepsilon \\ \textbf{draw} & \text{if } |A_{\text{eff}_{\text{in}}} - A_{\text{eff}_{\text{out}}}| \leq \varepsilon \quad (2) \\ \textbf{competitor} & \text{otherwise} \end{cases}$$

Here, $A_{\text{eff}}$ represents the effective amount received by the user, which will be defined in more detail shortly.

2. Defining a parity threshold ($\varepsilon$) to account for negligible differences in performance:

$$\varepsilon_{\text{USD}} = \begin{cases} \$1 & \text{if } V_{\text{in}} < \$10{,}000 \\ \$5 & \text{if } \$10{,}000 \leq V_{\text{in}} < \$100{,}000 \\ \$10 & \text{if } \$100{,}000 \leq V_{\text{in}} < \$500{,}000 \\ \$50 & \text{otherwise} \end{cases} \quad (3)$$

where:

$\varepsilon$     Parity threshold in USD

$V_{\text{in}}$     Transaction volume in USD (is equal to incoming tx amount of `src_token` in USD)

3. The effective amount ($A_{\text{eff}}$) represents the net value received by the user after accounting for transaction costs:

$$A_{\text{eff}} = \underbrace{\overbrace{A_{\text{dst}}^{\text{raw}}}^{\text{trace}} \times p_{\text{dst}}}_{\text{dst amount}} - \underbrace{\overbrace{G_{\text{used}}}^{\text{trace}} \times \overbrace{p_{\text{gas}}}^{\text{mined}} \times p_{\text{native}}}_{\text{tx cost}} \quad (4)$$

where:

$A_{\text{eff}}$     Effective amount to user in USD

$A_{\text{dst}}^{\text{raw}}$     Destination token amount to user parsed from transfers in traces

$p_{\text{dst}}$     Price of the destination token in USD

$G_{\text{used}}$     Gas used from trace

$p_{\text{gas}}$     Gas price from mined incoming transaction (for comparison flow with classics); Gas price from instant preset of outgoing transaction (for comparison flow with intents)

$p_{\text{native}}$     Price of the native token in USD

In the context of intent-based protocols such as 1inch Fusion and UniswapX, the conventional approach to gas price calculation is notably altered. Unlike traditional swap protocols where users directly incur gas costs, intent-based systems incorporate these expenses into the settlement price. This integration is facilitated by resolvers, who assume responsibility for gas payments as part of their operational model. Consequently, the effective amount ($A_{\text{eff}}$) calculation for intent-based transactions does not require a separate gas cost component. Instead, this cost is implicitly factored into the destination token amount ($A_{\text{dst}}^{\text{raw}}$) received by the user, reflecting a more holistic representation of the transaction's economic impact.

This framework allows DECS to compare transactions on a level playing field, taking into account both the amount of tokens received and the cost of executing the transaction.

Using these foundational metrics, additional performance indicators are derived:

$$\text{Uplift (\$)} = A_{\text{eff}_{\text{out}}} - A_{\text{eff}_{\text{in}}} \quad (5)$$

$$\text{Uplift (\%)} = \frac{A_{\text{eff}_{\text{out}}} - A_{\text{eff}_{\text{in}}}}{V_{\text{in}}} \quad (6)$$

$$\text{1inch winrate} = \frac{W_{\text{out}}}{W_{\text{in}}} \quad (7)$$

where:

$W_{\text{out}}$     quantity of transactions where winner is outgoing transaction

$W_{\text{in}}$     quantity of transactions where winner is incoming transaction

These metrics provide a comprehensive view of 1inch's performance relative to competitors, allowing us to quantify improvements in user outcomes across various transaction sizes and market conditions.

## V. COMPARISON RESULTS

### A. Classic mode

#### 1. *Ethereum*

| Winner | Comparisons | % won | p05 | Median | Mean | p95 |
|---|---|---|---|---|---|---|
| | | | | *Uplifts* | | |
| TOTAL | 534,540 | - | –$0.07 | $2.04 | $7.99 | $27.97 |
| 1inch | 378,547 | 71% | $1.14 | **$3.28** | $12.14 | $36.32 |
| Parity | 141,865 | 27% | –$0.33 | $0.55 | $0.52 | $0.98 |
| Competitors | 14,128 | 3% | –$1.12 | –$3.72 | –$28.12 | –$88.32 |

| | |
|---|---|
| Total volume of analysed transactions | $2,530,926,769 |
| Total uplift | $4,273,546 (0.17%) |
| Times 1inch is better than nearest competitor | **27x** |

Table 1: Benchmarks for classics - Ethereum, all buckets

Furtheron, we will refer to traditional DEX aggregation protocols as "classics". Over the course of 6 months (February to August), we conducted 534,540 comparisons on the Ethereum network, evaluating the performance of 1inch Classic against several major competitors. The results (*Table 1*) showed that 1inch offered better rates in 71% of transactions (378,547 cases), achieved parity in 27% of cases (141,865 cases), and was outperformed in 3% of comparisons (14,128 cases). The ratio of 1inch wins to competitor wins was approximately 27:1.

In cases where 1inch Classic won, the median uplift (5) was $3.28, with an average of $12.14. The uplift distribution for 1inch wins ranged from $1.14 at the 5th percentile to $36.32 at the 95th percentile. For transactions resulting in parity, the median uplift was $0.55, with an average of $0.52. The 5th percentile uplift for parity cases was −$0.33, while the 95th percentile was $0.98. In the relatively rare cases (3%) where competitors outperformed 1inch, the median uplift in favor of the competitors was $3.72, with an average of $28.12.

Overall, the total uplift across all comparisons amounted to $4,273,546, representing 0.17% of the total volume of analyzed transactions ($2,530,926,769).

**Breaking down the analysis by volume buckets:**

For transactions under $10,000 (*Table 2*), which comprised 502,613 of the analyzed trades, 1inch Classic won in 71% of cases (357,325 transactions). In these winning instances, the median uplift was $3.08, with an average uplift of $6.43. The uplift distribution for 1inch wins in this volume bucket ranged from $1.14 at the 5th percentile to $22.48 at the 95th percentile.

Parity was achieved in 26% of cases (132,529 transactions), with a median uplift of $0.54 and an average of $0.48. The 5th percentile uplift for parity cases was −$0.28, while the 95th percentile was $0.95. Competitors outperformed 1inch in 3% of comparisons (12,759 cases). In these instances, the median uplift in favor of the competitors was $3.21, with an average of $12.27. The 5th percentile uplift for competitor wins was $1.10, while the 95th percentile was $57.09.

The win-loss ratio in this category was approximately 28:1 in favor of 1inch. The total uplift across all comparisons in this bucket amounted to $2,204,571, representing 0.43% of the total volume of analyzed transactions ($512,825,509).

| | | | | Uplifts | | |
|---|---|---|---|---|---|---|
| **Winner** | **Comparisons** | **% won** | **p05** | **Median** | **Mean** | **p95** |
| TOTAL | 502,613 | - | -$0.01 | $1.97 | $4.39 | $18.22 |
| 1inch | 357,325 | 71% | $1.14 | **$3.08** | $6.43 | $22.48 |
| Parity | 132,529 | 26% | −$0.28 | $0.54 | $0.48 | $0.95 |
| Competitors | 12,759 | 3% | −$57.09 | −$3.21 | −$12.27 | −$1.10 |

| Volume of analysed transactions | $512,825,509 |
|---|---|
| Total uplift | $2,204,571 (0.43%) |
| Times 1inch is better than nearest competitor | **28x** |

Table 2: Benchmarks for classics - Ethereum, < $10k bucket

In the $10,000 to $100,000 range (*Table 3*), which included 28,756 transactions, 1inch Classic maintained a strong performance with a win rate of 67% (19,201 cases). For winning trades in this volume range, the median uplift was $41.43. The average uplift was $71.28. The uplift distribution showed a range with the 5th percentile at $9.56 and the 95th percentile at $208.33.

Parity was achieved in 29% of cases (8,348 transactions), with a median uplift of $0.96 and an average of $1.06. The 5th percentile uplift for parity cases was −$1.75, while the 95th percentile was $3.84. Competitors outperformed 1inch in 4% of comparisons (1,207 cases). In these instances, the median uplift in favor of the competitors was $27.71, with an average of $86.39. The 5th percentile uplift for competitor wins was $5.78, while the 95th percentile was $289.42. The win-loss ratio in this category was approximately 16:1 in favor of 1inch. The total uplift across all comparisons in this bucket amounted to $1,273,115.

| | | | | Uplifts | | |
|---|---|---|---|---|---|---|
| **Winner** | **Comparisons** | **% won** | **p05** | **Median** | **Mean** | **p95** |
| TOTAL | 28,756 | - | −$2.91 | $26.32 | $44.27 | $164.97 |
| 1inch | 19,201 | 67% | $9.56 | **$41.43** | $71.28 | $208.33 |
| Parity | 8,348 | 29% | −$1.75 | $0.96 | $1.06 | $3.84 |
| Competitors | 1,207 | 4% | $5.78 | −$27.60 | −$86.50 | −$289.42 |

| Volume of analysed transactions | $756,910,541 |
|---|---|
| Total uplift | $1,273,115 (0.17%) |
| Times 1inch is better than nearest competitor | **16x** |

Table 3: Benchmarks for classics - Ethereum, $10k-100k bucket

For transactions exceeding $100,000 (*Table 4*), which accounted for 3,171 high-value trades, 1inch Classic won in 64% of cases (2,021 transactions). In these winning instances, the median uplift was $230.75, with an average uplift of $460.38. The uplift distribution for 1inch wins in this volume bucket ranged from $20.45 at the 5th percentile to $1,680.70 at the 95th percentile. Parity was achieved in 31% of cases (988 transactions), with a median uplift of $0.89 and an average of $1.80. The 5th percentile uplift for parity cases was −$4.98, while the 95th percentile was $9.07.

| | | | | Uplifts | | |
|---|---|---|---|---|---|---|
| **Winner** | **Comparisons** | **% won** | **p05** | **Median** | **Mean** | **p95** |
| TOTAL | 3,171 | - | −$13.71 | $98.89 | $250.98 | $1,334.06 |
| 1inch | 2,021 | 64% | $20.45 | **$230.75** | $460.38 | $1,680.70 |
| Parity | 988 | 31% | −$4.98 | $0.89 | $1.80 | $9.07 |
| Competitors | 162 | 5% | −$11.32 | −$109.11 | −$841.66 | -$2,956.95 |

| Volume of analysed transactions | $1,261,190,720 |
|---|---|
| Total uplift | $795,860 (0.06%) |
| Times 1inch is better than nearest competitor | **12x** |

Table 4: Benchmarks for classics - Ethereum, > $100k bucket

In comparison, when competitors outperformed 1inch (162 cases, 5% of transactions), the median uplift in favor of the competitors was $109.11, with an average of $841.66. The uplift distribution for competitor wins ranged from $11.32 at the 5th percentile to −$2956.95 at the 95th percentile. The win-loss ratio in this category was approximately 12:1 in favor of 1inch Classic. The total uplift across all comparisons in this bucket amounted to $795,860, representing 0.06% of the total volume of analyzed transactions ($1,261,190,720).

Some key points:

1. 1inch Classic consistently provides better rates across all volume ranges, with win rates ranging from 64% to 71%.
2. The win-loss ratio strongly favors 1inch Classic, ranging from 12:1 to 28:1.
3. The magnitude of 1inch's relative outperformance increases with transaction size. For winning trades, the median uplift rises from $3.08 for transactions under $10,000 (3 bps) to $230.75 for transactions over $100,000 (23 bps).
4. In the under $10,000 range, which comprises the majority of trades (502,613), 1inch maintains a high win rate of 71% with a median uplift of $3.08.
5. For medium-sized transactions ($10,000 to $100,000), 1inch's win rate slightly decreases to 67%, but the median uplift increases substantially to $41.43.
6. In large transactions (over $100,000), 1inch's win rate further decreases to 64%, but the potential for significant outperformance increases, with a median uplift of $230.75 and an average uplift of $460.38.
7. The uplift distribution widens as transaction size increases, indicating greater variability in performance for larger trades.
8. When competitors outperform 1inch in the over $100,000 category, they do so by substantial margins, with a median uplift of $109.11 and an average of $841.66 in their favor.

## 2. *Arbitrum, Binance Smart Chain and Polygon*

| | | | Uplifts | | | |
|---|---|---|---|---|---|---|
| **Winner** | **Comparisons** | **% won** | **p05** | **Median** | **Mean** | **p95** |
| TOTAL | 472,264 | - | −$0.07 | $0.02 | $0.40 | $1.40 |
| 1inch | 30,950 | 7% | $1.06 | **$2.17** | $6.66 | $20.64 |
| Parity | 438,586 | 93% | −$0.07 | $0.02 | $0.08 | $0.51 |
| Competitors | 2,728 | 1% | −$1.05 | −$2.34 | −$20.20 | −$55.50 |

| | |
|---|---|
| Total volume of analysed transactions | $347,492,444 |
| Total uplift | $187,879 (0.05%) |
| Times 1inch is better than the nearest competitor | **11x** |

Table 5: Benchmarks for classics - non Ethereum, all buckets

Over the same 6-month period, DECS accumulated 472,264 comparisons across non-Ethereum networks, evaluating 1inch Classic against major competitors.

Across all volume buckets (*Table 5*), 1inch offered better rates in 7% of transactions (30,950 cases), achieved parity in 93% of cases (438,586 cases), and was outperformed in 1% of comparisons (2,728 cases). The ratio of 1inch wins to competitor wins was approximately 11:1.

In cases where 1inch Classic won, the median uplift was $2.17, with an average of $6.66. The uplift distribution for 1inch wins ranged from $1.06 at the 5th percentile to $20.64 at the 95th percentile. For transactions resulting in parity, the median uplift was $0.02, with an average of $0.08. The 5th percentile uplift for parity cases was −$0.07, while the 95th percentile was $0.51. In the rare cases (1%) where competition outperformed 1inch, the median uplift in favor of the competitors was $2.34, with an average of $20.27.

Overall, the total uplift across all comparisons amounted to $187,879, representing 0.05% of the total volume of analyzed transactions ($347,492,444).

**Breaking down the analysis by volume buckets.**

| | | | Uplifts | | | |
|---|---|---|---|---|---|---|
| **Winner** | **Comparisons** | **% won** | **p05** | **Median** | **Mean** | **p95** |
| TOTAL | 466,179 | - | −$0.06 | $0.02 | $0.26 | $1.30 |
| 1inch | 29,495 | 6% | $1.06 | **$2.05** | $4.42 | $13.80 |
| Parity | 434,205 | 93% | −$0.06 | $0.02 | $0.08 | $0.50 |
| Competitors | 2,479 | 1% | −$1.05 | −$2.03 | −$18.11 | −$54.34 |

| | |
|---|---|
| Volume of analysed transactions | $207,160,950 |
| Total uplift | $122,396 (0.06%) |
| Times 1inch is better than the nearest competitor | **12x** |

Table 6: Benchmarks for classics - non Ethereum, < $10k bucket

For transactions under $10,000 (*Table 6*), which comprised 466,179 of the analyzed trades, 1inch Classic won in 6% of cases (29,495 transactions). In these winning instances, the median uplift was $2.05, with an average uplift of $4.42. The uplift distribution for 1inch wins in this volume bucket ranged from $1.06 at the 5th percentile to $13.80 at the 95th percentile.

Parity was achieved in 93% of cases (434,205 transactions), with a median uplift of $0.02 and an average of $0.08. The 5th percentile uplift for parity cases was −$0.06, while the 95th percentile was $0.50.

Competitors outperformed 1inch in 1% of comparisons (2,479 cases). In these instances, the median uplift in favor of the competitors was $2.03, with an average of $18.11. The 5th percentile uplift for competitor wins was $1.05, while the 95th percentile was $54.34.

In the $10,000 to $100,000 range (*Table 7*), which included 5,998 transactions, 1inch Classic showed improved performance with a win rate of 24% (1,426 cases). For winning trades in this volume range, the median uplift was $25.18, with an average uplift of $46.74. The uplift distribution showed a range with the 5th percentile at $6.56 and the 95th percentile at $139.11.

|  |  |  | *Uplifts* | | | |
| --- | --- | --- | --- | --- | --- | --- |
| **Winner** | **Comparisons** | **% won** | **p05** | **Median** | **Mean** | **p95** |
| TOTAL | 5,998 | - | −$4.10 | $0.06 | $9.38 | $54.07 |
| 1inch | 1,426 | 24% | $6.56 | **$25.18** | $46.74 | $139.11 |
| Parity | 4,324 | 72% | −$2.55 | $0.02 | $0.00 | $2.42 |
| Competitors | 248 | 4% | -$5.18 | $8.77 | $41.80 | -$69.63 |

| | |
| --- | --- |
| Volume of analysed transactions | $127,208,915 |
| Total uplift | $56,280 (0.04%) |
| Times 1inch is better than the nearest competitor | **6x** |

Table 7: Benchmarks for classics - non Ethereum, $10k-100k bucket

Parity was achieved in 72% of cases (4,324 transactions), with a median uplift of $0.02. Competitors outperformed 1inch in 4% of comparisons (248 cases), with a median uplift of $8.77 in their favor.

|  |  |  | *Uplifts* | | | |
| --- | --- | --- | --- | --- | --- | --- |
| **Winner** | **Comparisons** | **% won** | **p05** | **Median** | **Mean** | **p95** |
| TOTAL | 87 | - | −$0.55 | $1.83 | $105.79 | $587.46 |
| 1inch | 29 | 33% | $12.24 | **$101.06** | $317.37 | $1,755.79 |
| Parity | 57 | 66% | −$3.73 | $0.64 | $1.14 | $6.89 |
| Competitors | 1 | 1% | −$65.23 | −$65.23 | −$65.23 | −$65.23 |

| | |
| --- | --- |
| Volume of analysed transactions | $13,122,580 |
| Total uplift | $9,203 (0.07%) |
| Times 1inch is better than the nearest competitor | **29x** |

Table 8: Benchmarks for classics - non Ethereum, > $100k bucket

For transactions exceeding $100,000 (*Table 8*), which accounted for 87 high-value trades, 1inch Classic won in 33% of cases (29 transactions). In these winning instances, the median uplift was $101.06, with an average uplift of $317.37. The uplift distribution for 1inch wins ranged from $12.24 at the 5th percentile to $1,755.79 at the 95th percentile.

Parity occurred in 66% of cases (57 transactions), while competitors outperformed 1inch in 1% of comparisons (1 case).

**Key observations and comparisons with Ethereum.**

1inch's overall win rate across all buckets on these networks (7%) is significantly lower than on Ethereum (71%), with a much higher parity rate (93% vs 27% on Ethereum). Despite the lower win rate, the win-loss ratio on these networks (12:1) remains favorable, though less pronounced than on Ethereum (27:1). For winning trades, the median and average uplifts are generally lower on these networks compared to Ethereum across all volume buckets.

Similar to Ethereum, the win rate and uplift magnitude increase with transaction size on these networks, but the increase is more pronounced. The win rate is 6% for transactions under $10,000, 24% for $10,000 to $100,000, and 33% for over $100,000, compared to 71%, 67%, and 64% respectively on Ethereum. In the over $100,000 category, while the win rate is lower than on Ethereum, the potential for significant outperformance remains high, with a median uplift of $101.06 and an average uplift of $317.37.

These networks show a much higher rate of parity across all volume buckets compared to Ethereum, indicating more frequent instances where 1inch matches competitor rates. Such behaviour can be explained by cheaper gas prices on these chains, which negatively impacts 1inch smart contract gas efficiency advantage. It's worth noting that the sample size for large transactions (over $100,000) is smaller on these networks (87 trades) compared to Ethereum (3,171 trades), which may affect the reliability of comparisons in this bucket.

## B. Intent mode

### 1. *1inch Classic vs 1inch Fusion*

*Ethereum*

| Bucket | n | p05 | p10 | p25 | p50 | Mean | p75 | p90 | p95 | WR$_{out}$ | Parity | LR$_{out}$ |
| --- | --- | --- | --- | --- | --- | --- | --- | --- | --- | --- | --- | --- |
| < $1k | 15,663 | 0.04% | 0.13% | 0.34% | 0.91% | 2.32% | 2.75% | 6.92% | 10.64% | 66.24% | 31.23% | 2.53% |
| $1k-10k | 10,565 | -0.17% | -0.02% | 0.03% | 0.08% | 0.05% | 0.17% | 0.31% | 0.47% | 73.6% | 17.2% | 9.2% |
| $10k-50k | 5,213 | -0.66% | -0.18% | -0.00% | 0.01% | -0.14% | 0.03% | 0.06% | 0.11% | 23.84% | 56.65% | 19.51% |
| $50k-100k | 1,469 | -0.60% | -0.26% | -0.02% | 0.00% | -0.07% | 0.01% | 0.02% | 0.03% | 21.99% | 46.9% | 31.11% |
| $100k-500k | 2,576 | -0.20% | -0.09% | -0.02% | -0.00% | -0.22% | 0.00% | 0.00% | 0.01% | 7.22% | 53.42% | 39.36% |
| $500k-1m | 390 | -0.23% | -0.15% | -0.07% | -0.01% | -0.10% | 0.00% | 0.00% | 0.01% | 5.64% | 40.77% | 53.59% |
| > $1m | 213 | -0.20% | -0.16% | -0.08% | -0.01% | -0.06% | 0.00% | 0.00% | 0.00% | 3.29% | 45.07% | 51.64% |
| Grand totals | 36,089 | -0.21% | -0.03% | 0.01% | 0.13% | 0.98% | 0.72% | 2.99% | 6.15% | 55.23% | 33.2% | 11.57% |

Table 9: 1inch Classic vs Fusion (Ethereum)

where:

WR$_{out}$      1inch Classic Win Rate

LR$_{out}$      1inch Classic Lose Rate

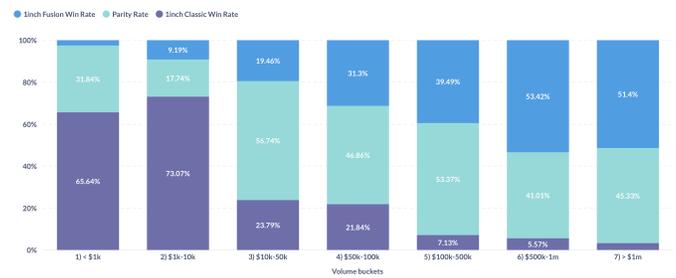

Figure 2: 1inch Classic vs 1inch Fusion winrates (Ethereum)

It is possible to further extend this comparison with more in depth manner by comparing 1inch own protocols between each other: 1inch Classic vs 1inch Fusion (*Table 9* and *Figure 2*).

For transactions under $1,000, which comprised 15,663 trades (43.40% of the total sample), 1inch Classic outper-

formed Fusion in 66.24% of cases. The median (p50) uplift was 0.91%, with an average uplift of 2.32%.

In the $1,000 to $10,000 range, encompassing 10,565 transactions (29.27% of the sample), 1inch Classic's performance improved further, winning in 73.60% of cases. The median uplift was 0.08%, and the average uplift was 0.05%.

However, a significant shift occurs for transactions between $10,000 and $50,000, covering 5,213 trades (14.44% of the sample). Here, 1inch Classic's win rate dropped to 23.84%, with Fusion outperforming in 19.51% of cases with 56.65% parity percent. The median uplift turned negative at 0.01%, with an average uplift of −0.14%.

This trend continues for larger transaction sizes:
- $50,000 to $100,000 (1,469 trades, 4.07% of the sample): Classic won 21.99% of the time, while Fusion won 31.11% with 46.90% parity.
- $100,000 to $500,000 (2,576 trades, 7.14% of the sample): Classic won only 7.22% of the time, with Fusion winning 39.36%.
- $500,000 to $1 million (390 trades, 1.08% of the sample): Classic's win rate further decreased to 5.64%, with Fusion winning 53.59%.
- Over $1 million (213 trades, 0.59% of the sample): Classic won in just 3.29% of cases, while Fusion won in 51.64%.

This data clearly shows that as order size increases, Fusion becomes increasingly effective compared to Classic. For transactions under $10,000, Classic maintains an advantage. Even though Classic provides better rates in retail cohort, raw comparisons do not reflect the cost of additional Fusion benefits: gasless swaps, MEV protection, smart swap execution (Dutch auction pricing) for the end user. Beyond this threshold, Fusion outperforms Classic.

Given prior comparisons in which 1inch Classic outperformed competitors across all volume ranges, it can be inferred that when Fusion outperforms Classic, it is likely also surpassing the competition. This is particularly significant for larger transactions where Fusion demonstrates a distinct superiority over Classic and, consequently, over other market participants.

*Arbitrum, Binance Smart Chain and Polygon*

| Bucket | n | p05 | p10 | p25 | p50 | Mean | p75 | p90 | p95 | WR$_{out}$ | Parity | LR$_{out}$ |
|---|---|---|---|---|---|---|---|---|---|---|---|---|
| < $1k | 49,704 | -0.14% | -0.06% | -0.01% | 0.01% | 0.36% | 0.16% | 1.05% | 2.51% | 0.27% | 99.23% | 0.49% |
| $1k-10k | 11,123 | -0.08% | -0.04% | -0.01% | -0.00% | -0.07% | 0.00% | 0.02% | 0.04% | 5.59% | 83.13% | 11.27% |
| $10k-50k | 2,336 | -0.04% | -0.02% | -0.01% | -0.00% | 0.02% | 0.00% | 0.02% | 0.04% | 7.28% | 85.57% | 7.15% |
| $50k-100k | 349 | -0.02% | -0.01% | -0.00% | -0.00% | 0.00% | 0.00% | 0.01% | 0.04% | 13.47% | 73.07% | 13.47% |
| $100k-500k | 201 | -0.06% | -0.02% | -0.00% | -0.00% | -0.02% | -0.00% | 0.01% | 0.02% | 11.94% | 71.64% | 16.42% |
| $500k-1m | 3 | N/A | N/A | N/A | N/A | N/A | N/A | N/A | N/A | N/A | N/A | N/A |
| > $1m | 1 | N/A | N/A | N/A | N/A | N/A | N/A | N/A | N/A | N/A | N/A | N/A |
| Grand totals | 63,717 | -0.12% | -0.05% | -0.01% | 0.00% | 0.27% | 0.08% | 0.74% | 1.89% | 1.57% | 95.69% | 2.74% |

Table 10: 1inch Classic vs 1inch Fusion winrates (non Ethereum)

Compared with Ethereum, a different dynamic is observed (*Table 10*):

For transactions under $1,000, 1inch Classic outperformed Fusion in only 0.27% of cases, with Fusion winning in 0.49% of cases. The vast majority (99.23%) of trades in this bucket resulted in parity. The median uplift is essentially insignificant, with an average uplift of 0.01%.

In the $1,000 to $10,000 range, encompassing 11,123 transactions (17.46% of the sample), 1inch Classic's performance improved slightly, winning in 5.59% of cases, while Fusion won in 11.27% of cases. Parity was achieved in 83.13% of trades. The median uplift was again insignificant, and the average uplift was 0.00%.

For transactions between $10,000 and $50,000, covering 2,336 trades (3.67% of the sample), 1inch Classic won in 7.28% of cases, while Fusion won in 7.15% of cases. Parity was achieved in 85.57% of trades.

In larger transaction buckets:
- $50,000 to $100,000 (349 trades, 0.55% of the sample): Classic won 13.47% of the time, Fusion won 13.47%, with 73.07% parity.
- $100,000 to $500,000 (201 trades, 0.32% of the sample): Classic won 11.94% of the time, Fusion won 16.42%, with 71.64% parity.

The comparison between 1inch Classic and Fusion on Arbitrum, Binance Smart Chain, and Polygon networks reveals significant differences from Ethereum. These chains show a much higher rate of parity between Classic and Fusion across all volume buckets, with lower win rates for both protocols compared to Ethereum. Unlike Ethereum, where Classic outperformed in smaller transactions, Fusion slightly edges out Classic in the under $10,000 range on these chains. The performance gap between Classic and Fusion is less pronounced as transaction sizes increase, contrasting with the clear trend seen on Ethereum. Average uplift values are generally smaller with minimal performance differences when there is a winner. The limited sample size for larger transactions (over $500,000) on these chains makes it challenging to draw definitive conclusions for high-value trades.

### 2. *1inch Classic vs UniswapX*

As established in *Section IV.B*, a methodical comparison of intent-based protocols requires first establishing baseline comparisons with 1inch Classic. Following the analysis of 1inch Fusion's performance relative to Classic, it is necessary to examine UniswapX's performance against the same baseline to enable subsequent indirect comparisons between intent-based solutions for common frame of reference in protocol evaluation.

| Bucket | n | p05 | p10 | p25 | p50 | Mean | p75 | p90 | p95 | WR$_{out}$ | Parity | LR$_{out}$ |
|---|---|---|---|---|---|---|---|---|---|---|---|---|
| < $1k | 36,655 | -0.32% | 0.01% | 0.41% | 0.91% | 1.30% | 1.93% | 3.33% | 4.40% | 74.74% | 19.84% | 5.42% |
| $1k-10k | 28,529 | -0.62% | -0.26% | 0.02% | 0.24% | 0.04% | 0.36% | 0.53% | 0.70% | 73.02% | 6.48% | 20.5% |
| $10k-50k | 11,023 | -1.10% | -0.42% | -0.07% | 0.16% | -0.66% | 0.23% | 0.28% | 0.33% | 62.46% | 9.3% | 28.24% |
| $50k-100k | 2,418 | -1.04% | -0.35% | -0.01% | 0.14% | -0.05% | 0.21% | 0.25% | 0.28% | 63.65% | 8.93% | 27.42% |
| $100k-500k | 1,919 | -0.83% | -0.44% | -0.02% | 0.08% | -0.07% | 0.20% | 0.24% | 0.29% | 61.07% | 7.82% | 31.11% |
| $500k-1m | 205 | -0.10% | -0.02% | 0.01% | 0.11% | 0.03% | 0.19% | 0.26% | 0.29% | 72.2% | 12.68% | 15.12% |
| > $1m | 140 | -0.02% | -0.01% | 0.00% | 0.10% | 0.09% | 0.16% | 0.20% | 0.23% | 75.71% | 8.57% | 15.71% |
| Grand totals | 80,889 | -0.57% | -0.20% | 0.09% | 0.31% | 0.51% | 0.84% | 2.12% | 3.18% | 71.8% | 13.04% | 15.16% |

Table 11: 1inch Classic vs UniswapX

During the period under review, 80,889 comparisons with UniswapX were accumulated (*Table 11*).

In the smallest transaction bucket (under $1,000), which represented 45.31% of the sample with 36,655 trades, 1inch Classic demonstrated strong dominance. It outperformed UniswapX in 74.74% of cases, with a median (p50) uplift of 0.91% and an average of 1.30%. The uplift distribution ranged from −0.32% (5th percentile) to 4.40% (95th percentile).

The $1,000 to $10,000 range, accounting for 35.27% of the sample (28,529 transactions), saw 1inch Classic maintain its edge with a 73.02% win rate. However, the magnitude of outperformance decreased, with the median uplift dropping to 0.24% and the average uplift to 0.04%. The uplift distribution narrowed, ranging from −0.62% at the 5th percentile to 0.70% at the 95th percentile.

A notable shift occurred in the $10,000 to $50,000 bucket (13.63% of the sample, 11,023 trades). Here, 1inch Classic's win rate dropped to 62.46%, with UniswapX outperforming in 28.24% of cases. The median uplift further decreased to 0.16%, while the average turned negative at −0.66%.

This trend continues for larger transaction sizes:
- $50,000 to $100,000 (2.99% of sample): 1inch Classic won 63.65% of cases, median uplift 0.14%
- $100,000 to $500,000 (2.37% of sample): Win rate dropped to 61.07%, median uplift 0.08%
- $500,000 to $1 million (0.25% of sample): Performance improved with 72.20% win rate, median uplift 0.11%
- Over $1 million (0.17% of sample): High 75.71% win rate maintained, median uplift 0.10%

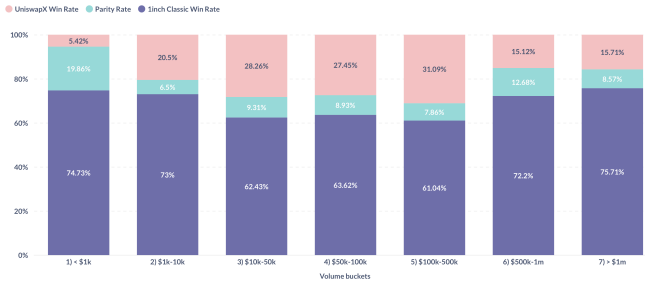

Figure 3: 1inch Classic vs UniswapX winrates (Ethereum)

The data suggests that 1inch Classic consistently outperforms UniswapX across all volume ranges, with varying degrees of advantage. The most significant outperformance occurs in smaller retail transactions (under $10,000), where 1inch Classic wins over 70% of the time with notable uplift percentages.

For medium-sized transactions ($10,000 to $500,000), while 1inch Classic still maintains a win rate above 50%, the margin of outperformance narrows.

For very large transactions (over $500,000), 1inch Classic's performance improves again, winning in over 70% of cases. However, the smaller sample size for these high-value transactions (361 total) limits the conclusions that can be drawn for this bucket.

The parity rate is highest in the lowest volume bucket (20.82% for transactions under $1,000) and generally decreases as transaction size increases, showing more pronounced performance differences in larger trades.

### 3. *Intent-based Protocols Comparison*

First of all, it should be noted that under the hood, 1inch Fusion utilizes 1inch Limit Order Protocol, which is the most gas-optimized among existing open limit protocols *[1]*. This enables 1inch Fusion to achieve significant gas savings when executing orders compared to other intent-based systems that rely on third-party solutions for trade execution.

Due to different underlying technologies between 1inch Classic and UniswapX, where first being an Aggregation protocol and the latter is an Intent-based protocol, for transparency, it is important to have same domain in this analysis.

By utilizing previous comparisons of 1inch Classic with 1inch Fusion, it is now possible to indirectly compare Fusion with UniswapX:

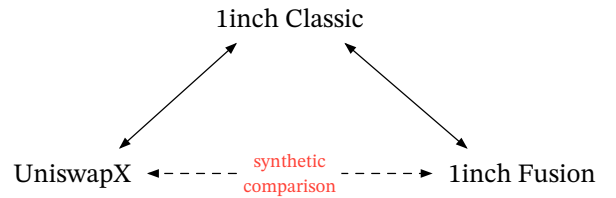

Figure 4: Comparison path

Given the methodological constraints outlined in *Section IV.B* that preclude direct comparison between 1inch Fusion and UniswapX, an alternative analytical approach becomes necessary. By utilizing 1inch Classic as a consistent baseline, it becomes feasible to compute the respective percentage uplift differentials for each protocol and subsequently conduct distributional analysis of these variations:

$$\text{Uplift}_{\text{Fusion}} (\%) \iff \text{Uplift}_{\text{UniswapX}} (\%) \qquad (8)$$

where

$$\text{Uplift (\%)} = \frac{A_{\text{eff}_{\text{out}}} - A_{\text{eff}_{\text{in}}}}{V_{\text{in}}} \quad (9)$$

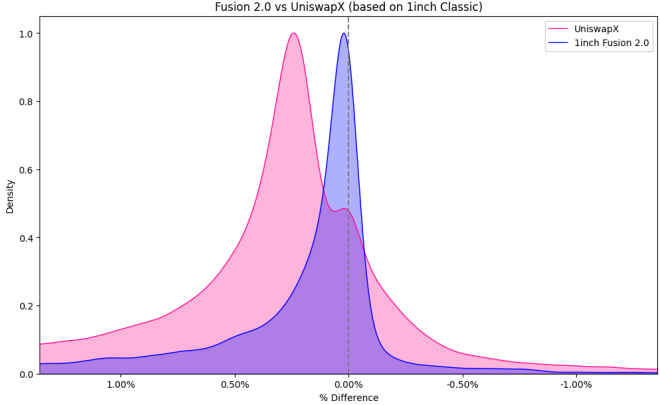

Figure 5: Intents' rate differences distributions

*Figure 5* illustrates the distribution of these uplift differences. Please note that x-axis is **inverted**, it represents the performance difference relative to 1inch Classic, with negative values indicating superior performance and positive values indicating underperformance relative to 1inch Classic (**right shift means better**). This inversion is necessitated by the requirement to evaluate incoming rather than outgoing transactions.

Building upon this baseline, pairwise comparisons are made between 1inch Classic and each intent-based protocol under examination. Given the directional nature of our research question - specifically investigating whether Fusion performs better than UniswapX - we employ one-tailed tests throughout our analysis.

A paired t-test was selected as the primary statistical tool due to the matched nature of our observations. While the normality assumption of parametric tests is traditionally important, the Central Limit Theorem *[7]* supports the use of t-tests with large sample sizes regardless of the underlying distribution. Additionally, we validated our findings using the non-parametric Mann-Whitney U test, which yielded consistent results, further supporting our approach.

Null Hypothesis ($H_0$):

$$H_0 : \mu_F \geq \mu_U \quad (10)$$

Alternative Hypothesis ($H_1$):

$$H_1 : \mu_F < \mu_U \quad (11)$$

where:

$\mu_F$    Mean difference in uplift for 1inch Fusion (relative to 1inch Classic)

$\mu_U$    Mean difference in uplift for UniswapX (relative to 1inch Classic)

| Bucket | n (G1,G2) | Mean (G1,G2) | CI (G1,G2) | CI diff | t-statistic | p-value | Cohen's d | Significant |
|---|---|---|---|---|---|---|---|---|
| Overall | 33,606 78,991 | 0.447% 0.576% | (0.437%, 0.458%) (0.569%, 0.584%) | (−0.142%, −0.116%) | −19.474 | < 1e−6 | −0.127 | + |

Table 12: Intents' rate differences distribution paired t-test results

There is a clear performance advantage for 1inch Fusion over UniswapX. With clear differences in centroids of these distributions (0.129%), Fusion demonstrates notable efficiency. Additionnaly, UniswapX distribution is wider and shifted further left, showing more variable and generally lower performance.

The statistical analysis of the uplift differences support the visual observations from *Figure 5*. A paired t-test was conducted to compare the two protocols, with results presented in *Table 12* (G1 being Fusion, G2 UniswapX). The test revealed a statistically significant difference (p-value < 1e-6) between Fusion and UniswapX, with a t-statistic of −19.474, thus, observed performance gap is statistically unlikely to be a result of random variation.

The mean uplift for Fusion (0.447%) was lower than that of UniswapX (0.576%), confirming Fusion's better overall rates relative to 1inch Classic. The 95% confidence interval for this difference (−0.142%, −0.116%) supports this conclusion.

Based on collected sample data, it is safe to say that 1inch Fusion rates are better than UniswapX by 0.116% to 0.142% with 95% confidence level. To better understand how this performance advantage varies across different trade sizes and to identify any potential trends based on swaps volume, it is necessary to go deeper and analyze whether this behavior persists across different swap amount sizes.

*Figure 6* presents a detailed breakdown of the performance comparison between 1inch Fusion and UniswapX across various transaction size buckets.

The distributions for both protocols in the smallest transaction buckets (0-100 and 100-500) show a relatively wide and overlapping range. The performance gap between Fusion and UniswapX for smaller trades shows inconsistency and variability. The Fusion distribution exhibits a minor shift to the right beginning at bucket 500, demonstrating an advantage in these lower volume ranges. As the transaction sizes progress from the range of $500-1k to $10k-50k, the distributions display a tendency to narrow and become more distinctly separated. The distribution of Fusion is con-sistently centered nearer to zero and demonstrates a more compact spread in comparison to UniswapX.

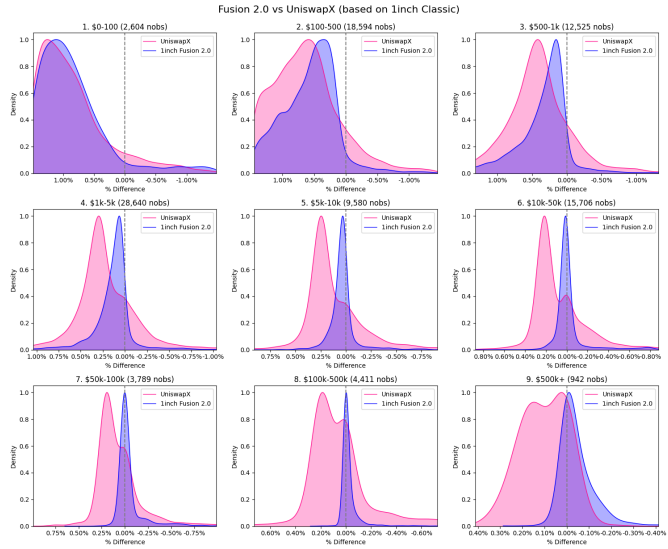

Figure 6: Intents' rate differences distributions by buckets

In the case of larger transactions, specifically those ranging from 50k to 100k and from 100k to 500k, the distinction between the two distributions is significantly more defined. The distribution of Fusion is characterized by a narrower range, with a central tendency that is closer to zero. In contrast, UniswapX shows a broader distribution with a distinct leftward shift. The data indicates that Fusion delivers improved and more reliable rates for high-value trades compared with UniswapX.

The performance gap achieves its maximum in the highest volume bucket, which is 500k and above. The distribution of Fusion is closely grouped around zero, signifying a high level of consistency in performance when compared to UniswapX. The distribution of UniswapX is characterized by a broader range and a shift further to the right. The significant disparity in the highest volume range highlights the efficiency of Fusion in managing large whale-level transactions.

The statistical analysis presented in *Table 13* quantifies the visual observations from *Figure 6*, providing rigorous support for the performance differences between 1inch Fusion and UniswapX across various transaction size buckets.

For statistical robustness, outliers beyond ±5% were excluded from the statistical tests to minimize the impact of extreme values on variance calculations.

Thresholds were determined through analysis of the data distribution, while excluding extreme outliers that often represent technical artifacts. For visualization clarity, the density plots display data within ±1.5% range, capturing the most relevant distribution characteristics while maintaining readability.

| Bucket | n (G1,G2) | Mean (G1,G2) | CI (G1,G2) | CI diff | t-statistic | p-value | Cohen's d | Significant |
|---|---|---|---|---|---|---|---|---|
| < $100 | 2,734<br>6,221 | 2.346%<br>2.265% | (2.291%, 2.402%)<br>(2.233%, 2.296%) | (0.018%, 0.146%) | 2.649 | 0.996 | 0.061 | − |
| $100–500 | 7,215<br>19,103 | 0.956%<br>1.171% | (0.932%, 0.979%)<br>(1.155%, 1.187%) | (−0.243%, −0.187%) | −14.173 | < 1e−6 | −0.196 | + |
| $500–1k | 3,319<br>9,916 | 0.373%<br>0.437% | (0.353%, 0.393%)<br>(0.423%, 0.451%) | (−0.088%, −0.040%) | −4.763 | < 1e−6 | −0.096 | + |
| $1k–5k | 7,844<br>21,275 | 0.132%<br>0.196% | (0.125%, 0.139%)<br>(0.189%, 0.203%) | (−0.074%, −0.054%) | −10.348 | < 1e−6 | −0.137 | + |
| $5k–10k | 2,686<br>7,031 | 0.011%<br>0.068% | (−0.000%, 0.022%)<br>(0.057%, 0.078%) | (−0.072%, −0.041%) | −5.989 | < 1e−6 | −0.136 | + |
| $10k–50k | 5,172<br>10,784 | −0.052%<br>0.013% | (−0.060%, −0.043%)<br>(0.004%, 0.022%) | (−0.077%, −0.053%) | −8.855 | < 1e−6 | −0.150 | + |
| $50k–100k | 1,466<br>2,406 | −0.098%<br>−0.006% | (−0.117%, −0.079%)<br>(−0.026%, 0.014%) | (−0.120%, −0.065%) | −6.165 | < 1e−6 | −0.204 | + |
| $100k–500k | 2,568<br>1,911 | −0.051%<br>−0.031% | (−0.060%, −0.041%)<br>(−0.054%, −0.007%) | (−0.045%, 0.005%) | −1.710 | 0.044 | −0.052 | + |
| > $500k | 602<br>344 | −0.062%<br>0.075% | (−0.076%, −0.049%)<br>(0.050%, 0.100%) | (−0.166%, −0.109%) | −10.449 | < 1e−6 | −0.706 | + |

Table 13: Intents' rate differences distributions paired t-test results by buckets

In the smallest bucket ($0-100), the at-test results show no statistically significant difference between the two protocols (p-value = 0.996). However, from the $100-500 bucket onwards, all comparisons show highly significant differences (p-values < 1e-6), confirming Fusion's consistent outperformance. This advantage persists in the highest volume bucket (>$500k), where Fusion maintains a significant edge of 0.109% to 0.166%, this represents a much larger dollar equivalent for these high-value transactions. Notably, the $100k-500k bucket shows a slightly smaller effect (Cohen's d = −0.052) compared to adjacent buckets, suggesting potential variability in performance advantages for very large transactions. However, the highest volume bucket (> $500k) demonstrates the most pronounced difference, with a Cohen's d of −0.706, confirming Fusion's particular efficiency in handling the largest transactions.

*Top token pairs (without stable pairs) comparisons*

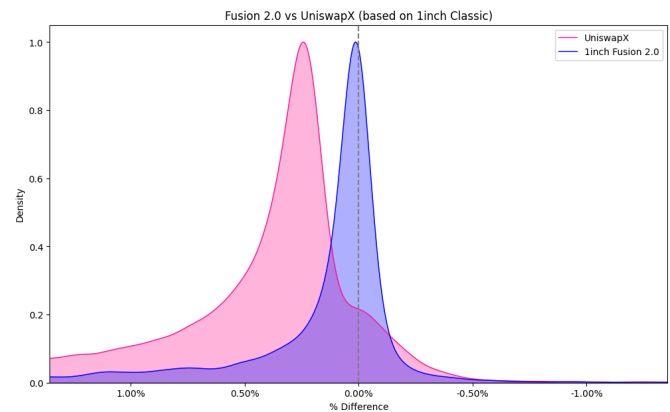

Figure 7: Intents' rate differences distributions (top token pairs, without stable pairs)

Current suggestion is that this advantage is possible due to partial order fills technology implemented in 1inch Fusion, contrary to UniswapX solution, however exact analysis of how partial fill impacts rates is not a subject of this paper.

Next, it is necessary to assess how UniswapX fees contribute to difference in rates, i.e. whether Fusion outperformance is solely connected to fees introduction by UniswapX. In order to conduct this analysis, it is important to have same-domain comparisons, therefore the top traded tokens were selected from the sample under consideration (ETH, DAI, USDC, USDT, WBTC, WETH, PEPE, USDe), included all permutations to construct traded pairs and excluded stable pairs as UniswapX does not take fees on stable swaps. *Figure 7* shows clear advantage of Fusion in this scenario

| Bucket | n (G1,G2) | Mean (G1,G2) | CI (G1,G2) | CI diff | t-statistic | p-value | Cohen's d | Significant |
|---|---|---|---|---|---|---|---|---|
| Overall | 12,014 50,048 | 0.385% 0.721% | (0.370%, 0.401%) (0.713%, 0.730%) | (−0.354%, −0.318%) | −35.857 | < 1e−6 | −0.364 | + |

Table 14: Intents' rate differences distributions paired t-test results (top token pairs)

According to *Table 14* it is confirmed that the difference in rates is statistically significant. With 95% confidence level, on top traded pairs (without stablecoin pairs), Fusion provides better rates (uplift of 0.318% - 0.354%). As Uniswap fee of 0.25% *[6]* is already included in these comparisons, and the observed difference (0.318% - 0.354%) exceeds this fee, it can be concluded that Fusion's performance advantage persists beyond the fee impact.

| Bucket | n (G1,G2) | Mean (G1,G2) | CI (G1,G2) | CI diff | t-statistic | p-value | Cohen's d | Significant |
|---|---|---|---|---|---|---|---|---|
| < $100 | 809 4,333 | 2.275% 2.293% | (2.173%, 2.377%) (2.258%, 2.327%) | (−0.126%, 0.090%) | −0.390 | 0.348 | −0.015 | − |
| $100–500 | 1,769 11,839 | 1.159% 1.346% | (1.106%, 1.212%) (1.328%, 1.364%) | (−0.243%, −0.131%) | −7.115 | < 1e−6 | −0.181 | + |
| $500–1k | 945 5,740 | 0.427% 0.606% | (0.393%, 0.461%) (0.594%, 0.619%) | (−0.216%, −0.143%) | −10.244 | < 1e−6 | −0.360 | + |
| $1k–5k | 2,510 13,125 | 0.159% 0.331% | (0.149%, 0.170%) (0.326%, 0.336%) | (−0.183%, −0.161%) | −27.347 | < 1e−6 | −0.596 | + |
| $5k–10k | 1,057 4,607 | 0.043% 0.206% | (0.036%, 0.049%) (0.201%, 0.212%) | (−0.172%, −0.155%) | −27.676 | < 1e−6 | −0.944 | + |
| $10k–50k | 2,492 7,278 | 0.003% 0.152% | (−0.001%, 0.007%) (0.148%, 0.157%) | (−0.156%, −0.144%) | −38.366 | < 1e−6 | −0.890 | + |
| $50k–100k | 730 1,609 | −0.028% 0.139% | (−0.036%, −0.021%) (0.130%, 0.147%) | (−0.179%, −0.155%) | −23.662 | < 1e−6 | −1.056 | + |
| $100k–500k | 1,368 1,294 | −0.044% 0.079% | (−0.050%, −0.038%) (0.063%, 0.094%) | (−0.140%, −0.106%) | −14.805 | < 1e−6 | −0.574 | + |
| > $500k | 334 223 | −0.099% 0.110% | (−0.119%, −0.078%) (0.072%, 0.147%) | (−0.251%, −0.166%) | −10.404 | < 1e−6 | −0.900 | + |

Table 15: Intents' rate differences distributions paired t-test results by buckets (top token pairs)

### Stable pairs comparisons

The next logical step is to compare 1inch performance on stable pairs (DAI, USDC, USDT, USDe).

In this scenario, a clear domain comparison is obtained as there are no fees on stable swaps on UniswapX.

| Bucket | n (G1,G2) | Mean (G1,G2) | CI (G1,G2) | CI diff | t-statistic | p-value | Cohen's d | Significant |
|---|---|---|---|---|---|---|---|---|
| Overall | 2,320 2,942 | 0.147% 0.296% | (0.126%, 0.168%) (0.272%, 0.321%) | (−0.181%, −0.117%) | −8.752 | < 1e−6 | −0.243 | + |

Table 16: Intents' rate differences distributions paired t-test results (stable pairs)

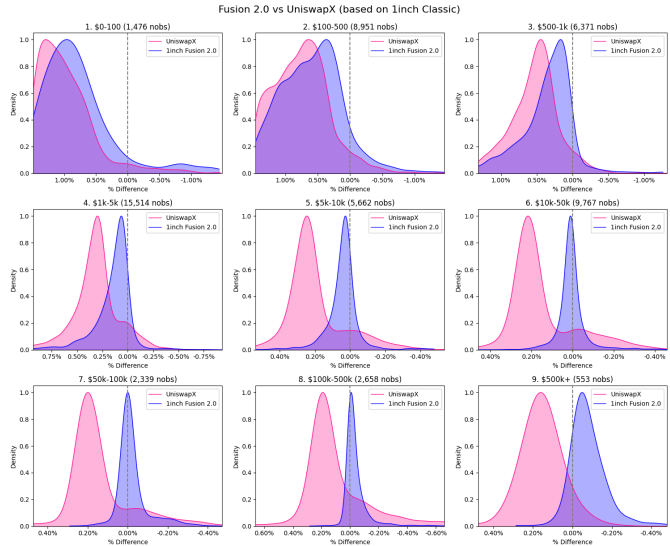

Figure 8: Intents' rate differences distributions by buckets (top token pairs)

*Figure 8* and *Table 15* confirm these findings.

With only one bucket failing to reject the null hypothesis (unable to conclude that Fusion performs better than UniswapX at the chosen significance level), the results achieved in *Table 14* demonstrate consistent performance advantages across most volume buckets. Please refer to *Table 15* for in-depth findings.

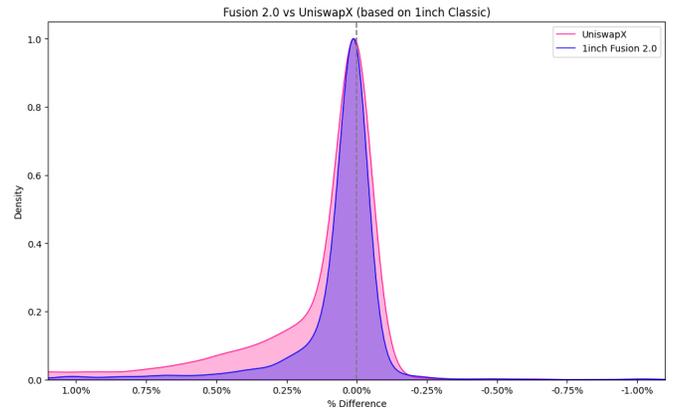

Figure 9: Intents' rate differences distributions (stable pairs)

*Table 16* confirms that Fusion outperforms UniswapX in this subset by 0.117% - 0.181% (95% confidence), such difference is crucial for stable coin swaps, especially in whale volume range. Notice the fatter left tail of UniswapX difference distribution in *Figure 9*. 8/9 buckets from *Table 17* show statistical signifiance with 90% confidence level. $\alpha = 0.1$ was chosen due to smaller sample compared with previous groups.

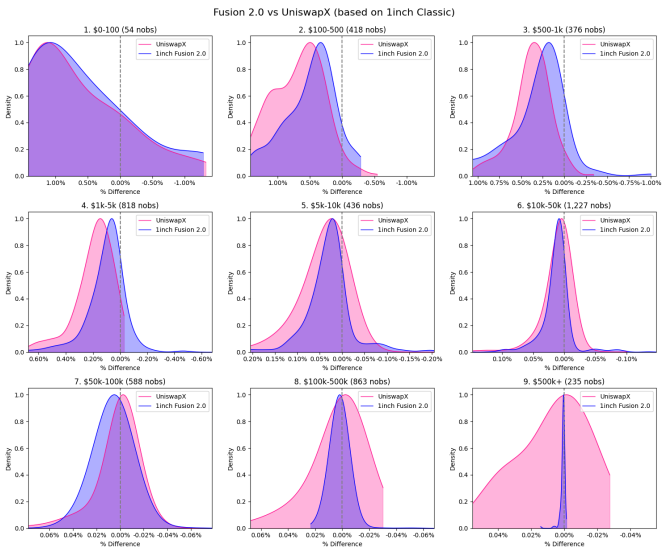

Figure 10: Intents' rate differences distributions by buckets (stable pairs)

| Bucket | n (G1,G2) | Mean (G1,G2) | CI (G1,G2) | CI diff | t-statistic | p-value | Cohen's d | Significant |
|---|---|---|---|---|---|---|---|---|
| < $100 | 58<br>121 | 1.865%<br>2.200% | (1.465%, 2.265%)<br>(1.994%, 2.407%) | (−0.781%, 0.111%) | −1.370 | 0.086 | −0.219 | + |
| $100–500 | 151<br>381 | 0.848%<br>1.018% | (0.729%, 0.967%)<br>(0.945%, 1.091%) | (−0.309%, −0.031%) | −2.031 | 0.021 | −0.195 | + |
| $500–1k | 122<br>262 | 0.364%<br>0.423% | (0.300%, 0.427%)<br>(0.383%, 0.462%) | (−0.133%, 0.015%) | −1.352 | 0.089 | −0.148 | + |
| $1k–5k | 434<br>384 | 0.106%<br>0.205% | (0.092%, 0.121%)<br>(0.189%, 0.221%) | (−0.120%, −0.077%) | −7.524 | < 1e−6 | −0.527 | + |
| $5k–10k | 233<br>203 | 0.029%<br>0.048% | (0.023%, 0.035%)<br>(0.036%, 0.061%) | (−0.033%, −0.006%) | −2.445 | 0.007 | −0.235 | + |
| $10k–50k | 534<br>693 | 0.013%<br>0.018% | (0.010%, 0.015%)<br>(0.015%, 0.021%) | (−0.010%, −0.002%) | −2.211 | 0.014 | −0.127 | + |
| $50k–100k | 210<br>378 | 0.004%<br>0.002% | (−0.001%, 0.010%)<br>(−0.001%, 0.005%) | (−0.004%, 0.009%) | 0.706 | 0.760 | 0.061 | − |
| $100k–500k | 464<br>399 | 0.001%<br>0.009% | (−0.001%, 0.003%)<br>(0.006%, 0.013%) | (−0.013%, −0.005%) | −3.600 | < 1e−6 | −0.246 | + |
| > $500k | 114<br>121 | 0.001%<br>0.012% | (0.001%, 0.001%)<br>(0.007%, 0.017%) | (−0.016%, −0.006%) | −3.467 | < 1e−6 | −0.453 | + |

Table 17: Intents' rate differences distributions paired t-test results by buckets (stable pairs, $\alpha = 0.1$)

# VI. Discussion

## A. Efficiency of Swaps in DeFi

Currently, there are two main approaches to implementing swaps in DeFi: **classics** (traditional DEX aggregation protocols) and **intent-based** solutions, which can be seen as an evolution of Limit Order Protocols. One of the key advantages of intent-based solutions (in addition to those outlined in section *Section II*) is the ability to partially fill orders. This allows for a broader range of liquidity to be utilized, including cold liquidity (described in the *Section II.A.1* ), which is especially important for executing large orders. However, classic solutions are currently still more effective for immediate trades where users need to execute transactions in the nearest block.

The efficiency of DeFi swaps depends on several key components:

1. **Gas efficiency of the protocol**. That is, how effectively the protocol works with specific exchange cases, how much gas will be spent on exchanging a specific pair for a given route, which pool families (Uniswap V2/V3, Curve, etc.) and which exchange route options it supports (the number of pools in the route, support for pools from different families within a single route, and so on).

2. **Efficiency of route construction**. This refers to the algorithm that searches for the most optimal route, balancing gas cost minimization with maximizing the received amount. An additional challenge here is the correct handling of tokens with transfer fees (FeeOn-Transfer tokens).

3. **Diversity of liquidity sources**. The greater the number of liquidity sources available for order execution, the higher the probability of finding the best rate and improving overall efficiency.

4. For intent-based solutions, additional factors include:
    - The number of **resolvers and their combined capabilities**. The aggregate liquidity coverage is important here, i.e. the volume that resolvers can use to execute orders.
    - Fair **competition among resolvers**, ensuring a level playing field for finding the best outcomes for users.
    - The effectiveness of the **price curve algorithm** *[5]*, which determines the price curve of order execution

Each of these factors requires detailed consideration to understand their impact on overall DeFi swap efficiency. As for the comparison of exchange protocols on efficiency in terms of gas costs, this is the most definite component of the overall efficiency of the exchange and 1inch has an open repository*[1]* for its comparison. Difficulties in objective comparison begin with the stage of route construction, with the choice of the moment of estimation closest to the real one, since the aspects described in section *Section II* are very significant.

Therefore, comparison systems like DECS, which incorporate the maximum number of efficiency factors, are essential for setting a benchmark and foundation for the objective comparison of DeFi swap products.

Another important consideration is the current rapid pace of DeFi development, which pushes projects to be highly dynamic in evolving their technologies. Many projects are actively enhancing their protocols, leading to dy-namic changes in their relative efficiency compared to other projects as improvements can simultaneously affect multiple key aspects.

Additionally, projects evolve asynchronously, adding complexity to the results of such analyses over a given period. Therefore, it is crucial to perform such comparisons over relatively short time intervals and in dynamic settings. However, shorter analysis periods may result in insufficient data. Even for rather broad 6-month period analyzed in this paper, some projects did not have enough data accumulated in particular volume buckets.

## B. For the End User

Given the complexity and multi-faceted nature of evaluating swap efficiency in DeFi, it is difficult for an average user without deep subject knowledge to objectively interpret the comparison results they encounter. Services like DECS can help bridge the gap between untrained users and their understanding of DeFi swap efficiency, thereby reducing the barrier to understanding such complex mechanics of the field.

It is also important to note that, beyond efficiency, several other factors are equally critical for end users when it comes to DeFi swaps:
- **Reliability and stability of expected outcomes**. Users should be provided with the statistically most favorable swap even under basic settings.
- **Transparency**. To provide a transparent, understandable description of mechanics and correct interpretation of the swap results, as well as for intent-based solutions to ensure the decentralization of processes and conditions for fair competition between solvers.
- **Security**. This includes the presence and quality of audits, code transparency, adherence to common security standards, and more.
- **Legal aspects**. Availability and improvement of systems to ensure control over the legality of the actions of the swap participants.

All of these factors, combined with efficiency, influence the overall objectivity of choosing a swap product.

## C. Directions for Further Development

1. **Improvement of analysis methodology**. To further advance comparison systems like DECS, it is crucial to continually refine the methodology used to account for new factors and more accurately assess swap efficiency.
2. **Increasing transparency**. It is important to develop tools that are accessible to third parties, allowing them to independently verify the results of systems like DECS and reproduce them. Such initiatives will strengthen trust in the results and contribute to the growth of the DeFi ecosystem.
3. **Expanding technical capabilities**. This includes the integration of new contracts and solutions, expanding the list of projects and blockchains on which comparisons are carried out. Also, in order to maximize transparency and correctness of comparisons, it is advisable to offer projects participation in open comparative assessments:
    - By providing access to their best APIs and information about which block the estimation was made on for a specific swap in classic solutions.
    - By offering information on executed orders for intent-based solutions.
4. **Dynamic analysis over time**. To better reflect the changes being implemented by projects, it is necessary to strive for dynamic evaluation of swap projects efficiency. This requires empirical selection of time intervals to ensure sufficient data volume for comparison.
5. **Improving accessibility for end users**. It is important to continue working on making the results of analysis understandable and useful even for untrained users, creating a "bridge" between objective metrics and simple interpretation for a wide range of DeFi participants.

## VII. CONCLUSION

### A. Final thoughts on the efficiency of DeFi exchanges

Due to the problems identified in the *Section VI*, the empirical results of this research are important for the DeFi ecosystem. They provide the clearest picture of the effectiveness of decentralized exchange protocols and confirm the technological advantages of the 1inch architecture. The stable superiority of 1inch in various exchange approaches, partial value exchanges and various blockchains demonstrates the comprehensive effectiveness of 1inch.

The analysis of 1inch Classic in comparison with intent-based solutions provides valuable information about the development of DeFi protocols. The emergence of intent-based solutions, especially their high performance when conducting large amount swaps, indicates a possible change in the paradigm of DeFi exchanges. This trend highlights the importance of constant innovation and adaptation to maintain market leadership.

A comparative analysis of 1inch Fusion and UniswapX revealed a statistically significant superiority of Fusion, especially in large exchanges, which suggests the fundamental importance of "partial fills" technology for intent-based solutions. These results indicate that leading efficiency is more related to fundamental algorithmic advantages, and much less to the commission structure.

The results of this study collectively highlight the crucial role of integrated efficiency analysis in protocol design and optimization. They provide useful information for further development, improvement and market positioning. In the face of fierce competition, services such as DECS are indispensable for improving all aspects of the effectiveness of DeFi swap. The ability to make detailed real-time comparisons for different market conditions, trading pairs, and trade volumes provides invaluable information for developers. This allows to identify specific areas for improvement and helps adapt solutions to meet the diverse needs of users on different networks.

## VIII. Acknowledgments

The authors would like to thank Mikhail Melnik and Anton Bukov for their insightful comments and suggestions that helped improve the quality of this paper.

## IX. Disclosure Statement

The authors are affiliated with 1inch. To ensure objectivity, this study employs standardized statistical methodologies, provides complete documentation of all analytical procedures, and makes the underlying data publicly available for independent verification.

## X. Appendix

For the benefit of readers seeking a more comprehensive analysis, additional data has been included in appendix that detail 1inch Classic's performance against all major competitors across various blockchain networks. These supplementary visualizations were omitted from the main text to maintain concision and focus.

The histograms presented herein illustrate the distribution of rate differences between 1inch Classic and each competitor. The x-axis represents the percentage difference in rates, with positive values indicating superior performance by 1inch Classic.

For visualization clarity, the histograms display data within ±5% range, capturing the most relevant distribution characteristics while maintaining readability.

To quantify the significance of these differences, statistical tests were conducted examining whether the proportion of positive differences across all buckets exceeds 0.50, thereby demonstrating a statistically significant advantage for 1inch Classic.

### 1. *Ethereum*

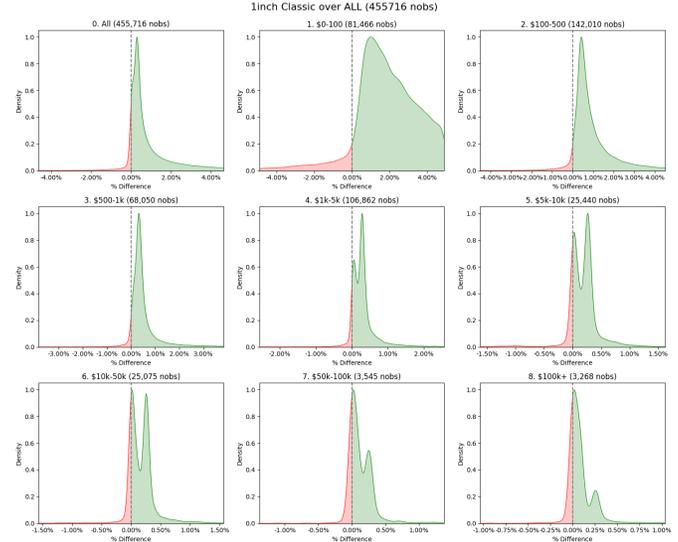

Figure 11: 1inch Classic vs All

| Competitor | n | Mean | CI | Proportion | z-statistic | P-value | Significant |
|---|---|---|---|---|---|---|---|
| All | 455,716 | 0.8063% | (0.8030%, 0.8095%) | 0.9458 | 601.8362 | < 1e-6 | + |

Table 18: All competitors test results

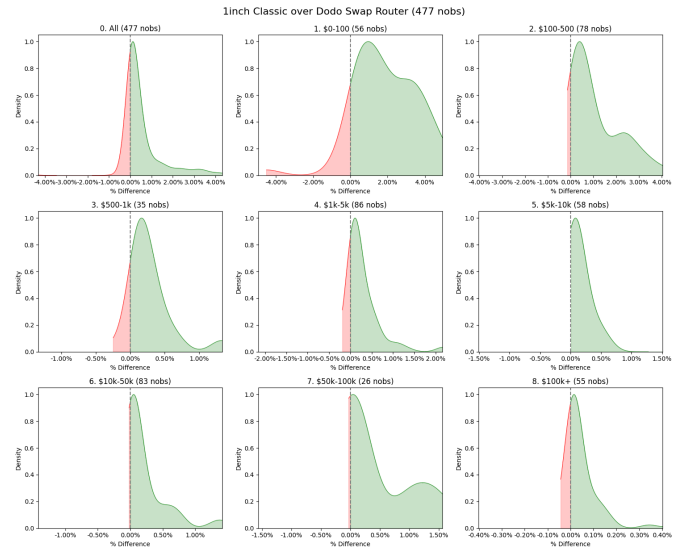

Figure 12: 1inch Classic vs Dodo

| Competitor | n | Mean | CI | Proportion | z-statistic | P-value | Significant |
|---|---|---|---|---|---|---|---|
| Dodo Swap | 477 | 0.5585% | (0.4699%, 0.6471%) | 0.9644 | 20.2836 | < 1e-6 | + |

Table 19: Dodo test results

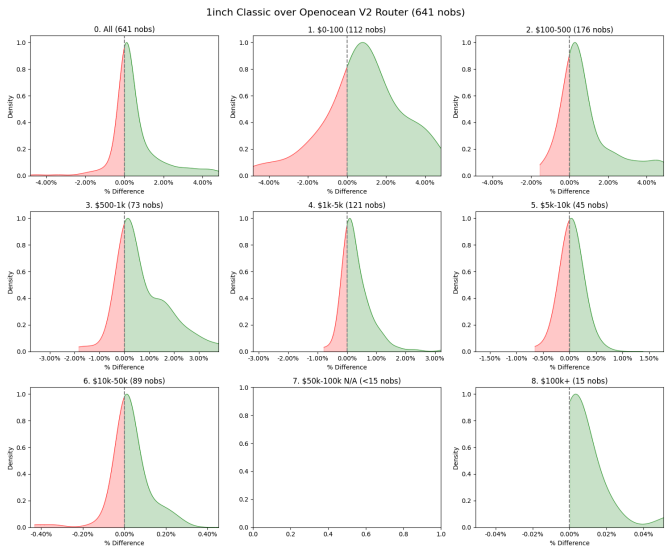

Figure 13: 1inch Classic vs Open Ocean

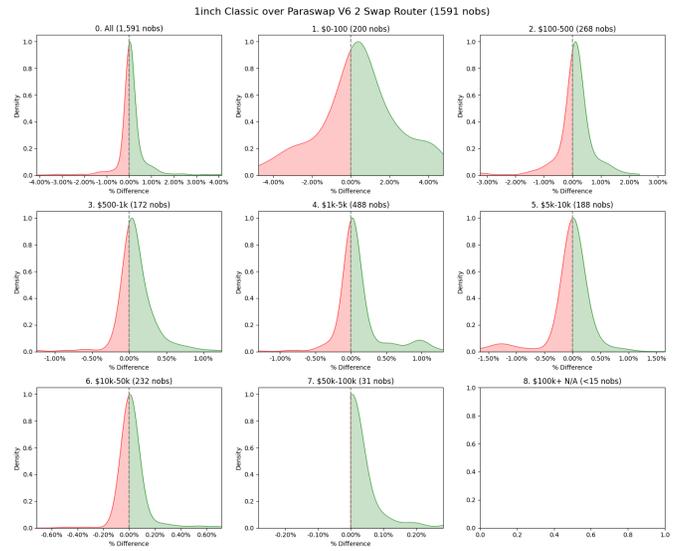

Figure 15: 1inch Classic vs Paraswap V6

| Competitor | n | Mean | CI | Proportion | z-statistic | P-value | Significant |
|---|---|---|---|---|---|---|---|
| Openocean V2 | 641 | 0.5618% | (0.4624%, 0.6611%) | 0.8097 | 15.6806 | < 1e-6 | + |

Table 20: Open Ocean test results

| Competitor | n | Mean | CI | Proportion | z-statistic | P-value | Significant |
|---|---|---|---|---|---|---|---|
| Paraswap V6 | 1,591 | 0.1067% | (0.0646%, 0.1488%) | 0.7178 | 17.3739 | < 1e-6 | + |

Table 22: Paraswap V6 test results

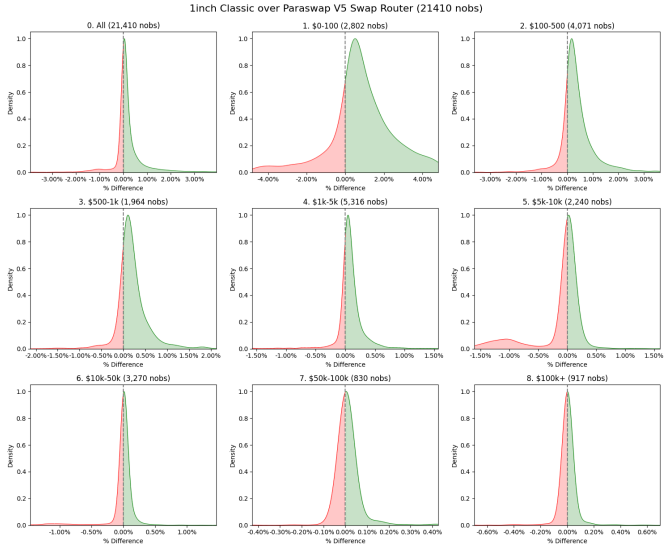

Figure 14: 1inch Classic vs Paraswap V5

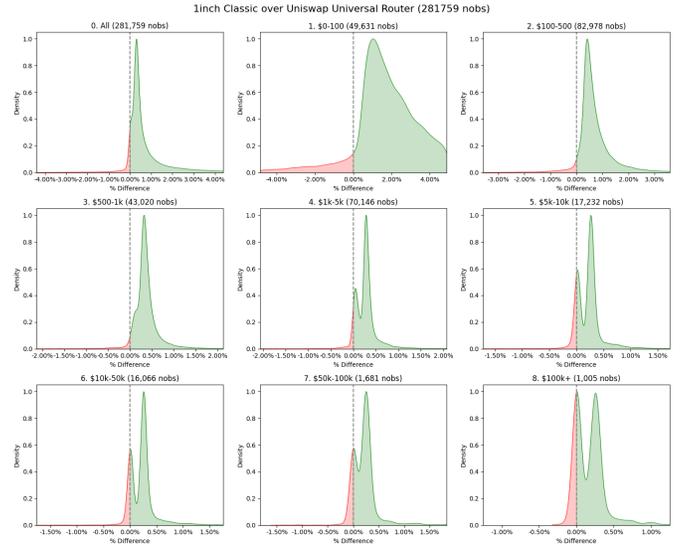

Figure 16: 1inch Classic vs Uniswap Router

| Competitor | n | Mean | CI | Proportion | z-statistic | P-value | Significant |
|---|---|---|---|---|---|---|---|
| Paraswap V5 | 21,410 | 0.2095% | (0.1987%, 0.2204%) | 0.8129 | 91.5791 | < 1e-6 | + |

Table 21: Paraswap V5 test results

| Competitor | n | Mean | CI | Proportion | z-statistic | P-value | Significant |
|---|---|---|---|---|---|---|---|
| Uniswap Router | 281,759 | 0.6667% | (0.6631%, 0.6704%) | 0.9493 | 476.9449 | < 1e-6 | + |

Table 23: Uniswap Router test results

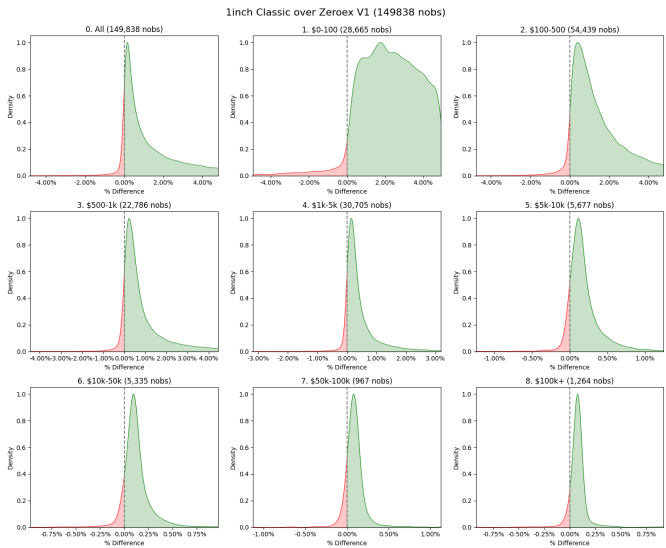

Figure 17: 1inch Classic vs Zeroex (Øx protocol)

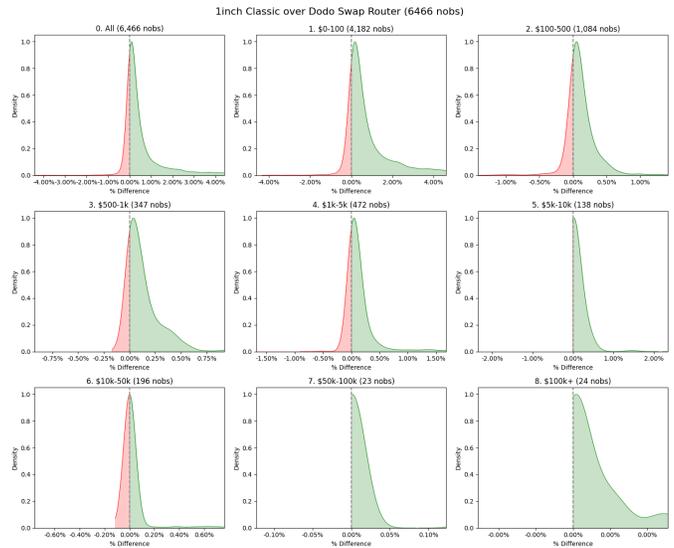

Figure 19: 1inch Classic vs Dodo

| Competitor | n | Mean | CI | Proportion | z-statistic | P-value | Significant |
|---|---|---|---|---|---|---|---|
| Zeroex V1 | 149,838 | 1.1631% | (1.1564%, 1.1698%) | 0.9611 | 356.9720 | < 1e-6 | + |

Table 24: Zeroex (Øx protocol) test results

| Competitor | n | Mean | CI | Proportion | z-statistic | P-value | Significant |
|---|---|---|---|---|---|---|---|
| Dodo Swap | 6,466 | 0.5658% | (0.5417%, 0.5900%) | 0.8356 | 53.9724 | < 1e-6 | + |

Table 26: Dodo test results

## 2. *Arbitrum, Binance Smart Chain and Polygon*

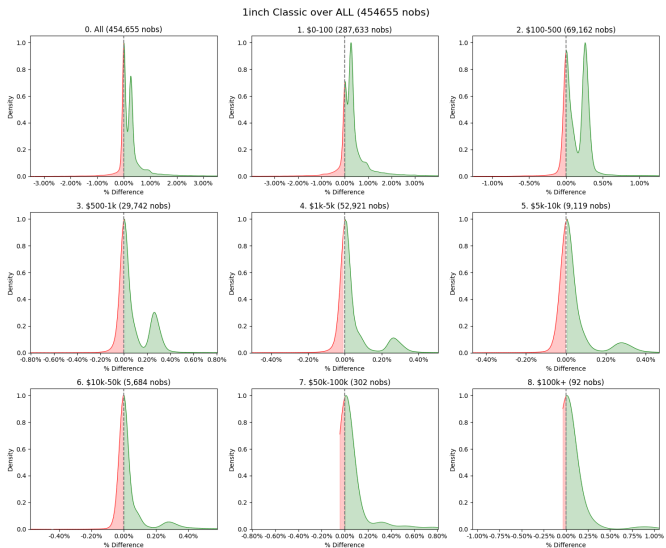

Figure 18: 1inch Classic vs All

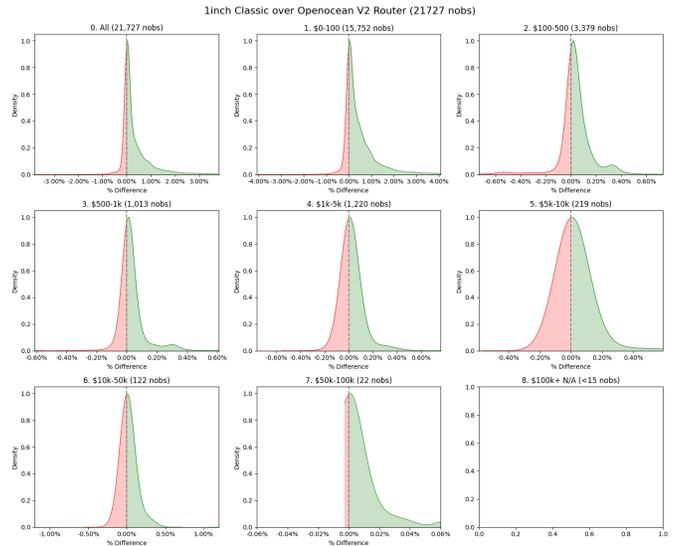

Figure 20: 1inch Classic vs Open Ocean

| Competitor | n | Mean | CI | Proportion | z-statistic | P-value | Significant |
|---|---|---|---|---|---|---|---|
| Openocean V2 | 21,727 | 0.3712% | (0.3613%, 0.3811%) | 0.7107 | 62.1096 | < 1e-6 | + |

Table 27: Open Ocean test results

| Competitor | n | Mean | CI | Proportion | z-statistic | P-value | Significant |
|---|---|---|---|---|---|---|---|
| All | 454,655 | 0.2674% | (0.2654%, 0.2693%) | 0.7461 | 331.9313 | < 1e-6 | + |

Table 25: All competitors test results

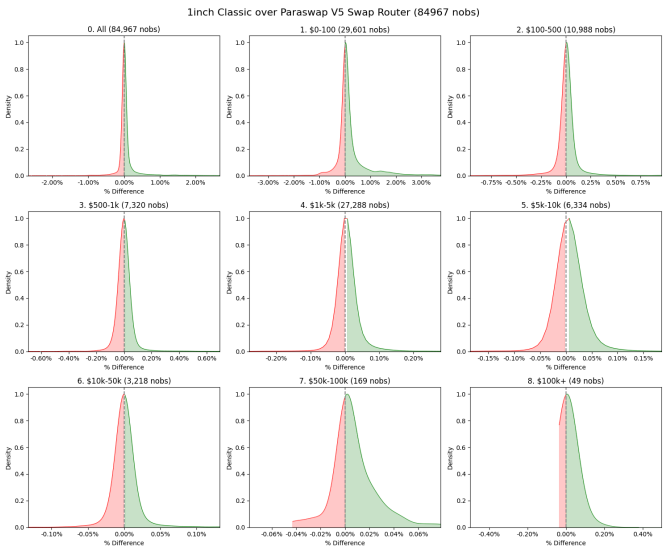

Figure 21: 1inch Classic vs Paraswap V5

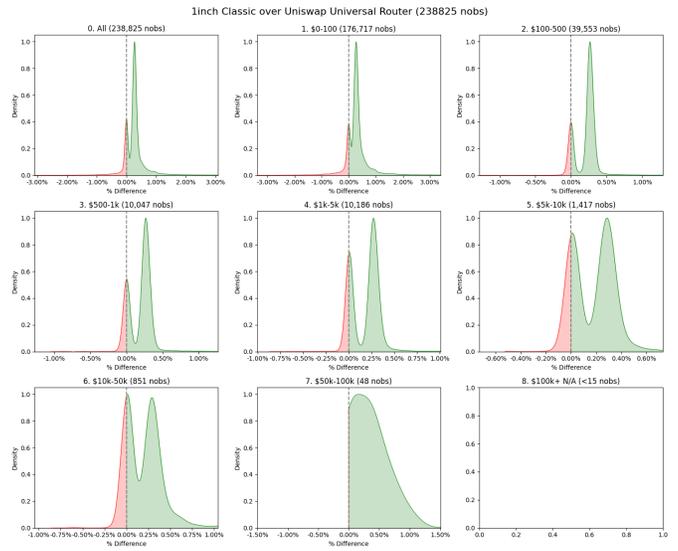

Figure 23: 1inch Classic vs Uniswap Router

| Competitor | n | Mean | CI | Proportion | z-statistic | P-value | Significant |
|---|---|---|---|---|---|---|---|
| Paraswap V5 | 84,967 | 0.0937% | (0.0903%, 0.0970%) | **0.5896** | 52.2520 | < 1e-6 | + |

Table 28: Paraswap V5 test results

| Competitor | n | Mean | CI | Proportion | z-statistic | P-value | Significant |
|---|---|---|---|---|---|---|---|
| Uniswap | 238825 | 0.2871% | (0.2848%, 0.2894%) | **0.7976** | 290.8815 | < 1e-6 | + |

Table 30: Uniswap Router test results

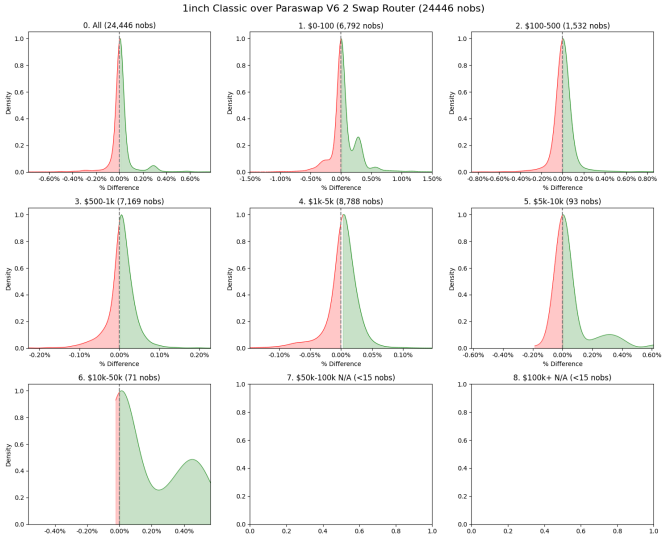

Figure 22: 1inch Classic vs Paraswap V6

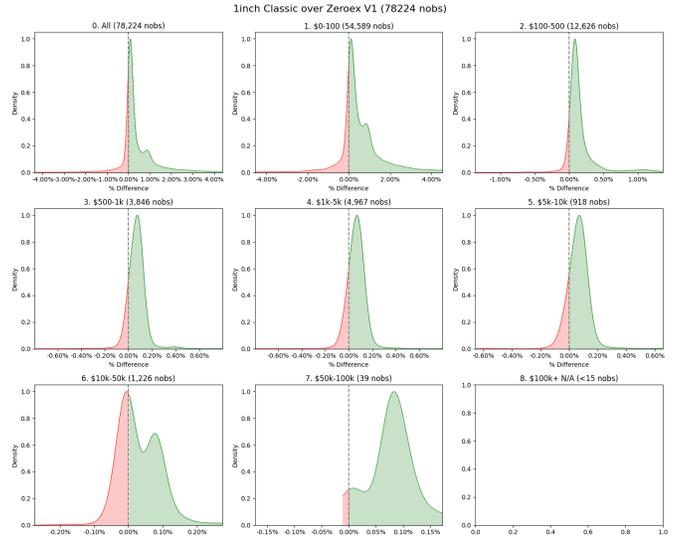

Figure 24: 1inch Classic vs Zeroex (Øx protocol)

| Competitor | n | Mean | CI | Proportion | z-statistic | P-value | Significant |
|---|---|---|---|---|---|---|---|
| Paraswap V6 | 24,446 | 0.0220% | (0.0195%, 0.0245%) | **0.5410** | 12.8172 | < 1e-6 | + |

Table 29: Paraswap V6 test results

| Competitor | n | Mean | CI | Proportion | z-statistic | P-value | Significant |
|---|---|---|---|---|---|---|---|
| Zeroex V1 | 78,224 | 0.4190% | (0.4120%, 0.4261%) | **0.8256** | 182.1045 | < 1e-6 | + |

Table 31: Zeroex (Øx protocol) test results

Figure 25: 1inch Classic vs Competitors in Ethereum - detailed table

Figure 26: 1inch Classic vs Competitors in Binance Smart Chain, Arbitrum and Polygon - detailed table

# References

[1] 1inch. Gas-Comparison. *GitHub*.

[2] 1inch. 1inch crushed the competition by 16%. (2024) *1inch Blog*.

[3] 1inch. DECS Whitepaper Data. (2024) *GiHub*.

[4] Guillermo Angeris and Tarun Chitra. Automated Market Making: Theory and Applications. (2020) :30–47.

[5] Anton Bukov, Sergej Kunz, Mikhail Melnik, Gleb Alekseev, Matt Snow, and Xenia Shape. 1inch Fusion+: Intent-based Atomic cross-chain swaps. (2024) *1inch Blog*.

[6] Uniswap Labs. What are Uniswap Labs' fees?.

[7] Dieter Rasch and Volker Guiard. The robustness of parametric statistical methods. (2004) *Psychology Science*, 462):175–208.

[8] Sam Werner, Daniel Perez, Lewis Gudgeon, and Ariah Klages-Mundt. SoK: Decentralized Finance (DeFi). (2021) :1–19.